\documentclass[pra,aps,eqsecnum,showpacs,twocolumn,superscriptaddress]{revtex4-1}
\usepackage{multirow}
\usepackage{amsmath}
\usepackage{amssymb}
\usepackage{latexsym}
\usepackage{MnSymbol}
\usepackage{color}
\usepackage{hyperref}
\usepackage{float}
\usepackage{graphicx}
\usepackage{bm}
\newtheorem{theorem}{Theorem}
\newtheorem{lemma}{Lemma}

\begin{document}
\newcommand{\ket} [1] {\vert #1 \rangle}
\newcommand{\bs} [1] {\boldsymbol{#1}}
\newcommand{\bra} [1] {\langle #1 \vert}
\newcommand{\braket}[2]{\langle #1 | #2 \rangle}
\newcommand{\proj}[1]{\ket{#1}\bra{#1}}
\newcommand{\mean}[1]{\langle #1 \rangle}
\newcommand{\opnorm}[1]{|\!|\!|#1|\!|\!|}
\newcommand{\hr}[1]{\hat{\rho}_{#1}}
\newcommand{\hatr}[1]{\hat{\rho}^{#1}}
\newcommand{\tr}[1]{{\rm Tr}\,\hatr{#1}}

\title{Divergence-free approach for obtaining decompositions of quantum-optical processes}
\author{K. K.  Sabapathy}
\affiliation{Centre for Quantum Information and Communication, Ecole polytechnique de Bruxelles, CP 165, Universit\'{e} libre de Bruxelles, 1050 Brussels, Belgium.}

\author{J. S.  Ivan}
\affiliation{Department of Physics,
Indian Institute of Space Science and Technology, Trivandrum 695 547, India.}

\author{R. Garc\'{i}a-Patr\'{o}n}
\affiliation{Centre for Quantum Information and Communication, Ecole polytechnique de Bruxelles, CP 165, Universit\'{e} libre de Bruxelles, 1050 Brussels, Belgium.}

\author{R. Simon}
\affiliation{Optics \& Quantum Information Group, The Institute of Mathematical
  Sciences, HBNI, C.I.T Campus, Tharamani, Chennai 600 113, India.}

\begin{abstract}
  Operator-sum representations of quantum channels can be obtained by applying the channel to one subsystem of a maximally entangled state and deploying the channel-state isomorphism. However, for continuous-variable systems, such schemes contain natural divergences since the maximally entangled state is ill-defined. We introduce a method that avoids such divergences by utilizing finitely entangled (squeezed) states and then taking the limit of arbitrary large squeezing. Using this method we 
derive an operator-sum representation for all single-mode bosonic Gaussian channels where a unique feature is that both quantum-limited and noisy channels are treated on an equal footing. This technique facilitates a proof that the rank-one Kraus decomposition for Gaussian channels at its respective entanglement-breaking thresholds, obtained in the overcomplete coherent state basis, is unique. The methods could have applications to simulation of continuous-variable channels.
\end{abstract}

\maketitle
\section{Introduction}
A quantum channel is a quantum process that describes valid state transformations of a given system. Quantum channels are mathematically described by  completely positive and trace preserving maps. Every quantum channel has a (non-unique) Stinespring\,\cite{stinespring} or unitary dilation, where the channel action on the system is described through a unitary interaction with an appended system that is subsequently discarded.

It is well-known that any quantum channel can be decomposed into what is known as an operator-sum representation or Kraus decomposition\,\cite{ecg61,kraus70}. In cases where the unitary dilation is known, the Kraus decomposition can be obtained in a straight-forward manner by evaluating suitable matrix elements of the unitary. However, if the channel is described by its action on states, an alternative method exists to obtain a Kraus decomposition that  uses the Choi-Jamiolkowski channel-state isomorphism\,\cite{choi75,jamiolkowski}. For finite-dimensional systems, the channel is applied to one subsystem of a maximally entangled state. From any pure state decomposition of the resulting state one obtains a corresponding set of Kraus operators.

The transition from finite  to infinite dimensions with respect to the Choi-Jamiolkowski isomorphism
is a nontrivial problem that needs a detailed discussion\,\cite{giedke02,holevo11,kiukas}. The main reason being that the notion of a `maximally entangled' state is `ill-defined' for continuous-variable systems.

To circumvent such technicalities and divergences, we devise a procedure which
makes effective use of the two-mode squeezed vacuum state \begin{align}
  |\Psi_{r}
  \rangle =  {\rm sech} \,r\, \sum_{j=0}^{\infty}
  (\tanh \, r)^j |j\,j\rangle, \end{align}
and techniques of phase-space quantum information theory.
The idea is to evaluate a pure state decomposition of the bipartite state $(\Phi \otimes {1\!\!1}) |\Psi_{r}
 \rangle \langle \Psi_{r} |$ and then take the limit $r \to \infty$ from which the Kraus operators of the channel $\Phi$ are obtained.
 We demonstrate our method by obtaining a Kraus decomposition for all single-mode bosonic Gaussian channels in the Fock and  coherent state basis.

Gaussian states and Gaussian channels have received considerable interest  recently in view of its applications to quantum information processing\,\cite{onea,oneb,onec,raulrmp}.  Also, bosonic Gaussian channels arise naturally in the description of many optical systems such as light transmission through optical fibers and amplifiers of optical signals \cite{twoa,twob,twoc,sm3}. The general structure of bosonic Gaussian channels were presented in \cite{cp3,eisert05}, entanglement-breaking Gaussian channels were characterized in \cite{holevo05,holevo08}, nonclassicality-breaking Gaussian channels were studied in \,\cite{gaussch,ncb,pon,scaling,nbm}, and the canonical forms of  single-mode Gaussian channels were detailed in \cite{holevo07a,holevo07b}. 

The operator-sum representations for all single-mode bosonic Gaussian channels
were obtained using the Stinespring dilations,  and its implications were studied in great detail in \cite{gaussch}. This Kraus decomposition
has proven particularly useful in  the demonstration of robustness of non-Gaussian entanglement over Gaussian entanglement against noisy environments \cite{robust}, apart from other information-theoretic applications.

Our new approach of obtaining Kraus operators of bosonic Gaussian channels allows us to
prove that the  operator-sum representation of entanglement-breaking Gaussian channels in terms of a
measurement-preparation in the over-complete coherent state basis is unique at its respective entanglement-breaking thresholds. The existence of a unique non-countable rank-one operator-sum representation
of entanglement-breaking channels of infinite dimensional systems was shown with the help of a  mathematical example in \cite{holevo05}.
Our result shows that this property is not a mathematical curiosity but is a natural occurrence among physically relevant channels that are widely used to model optical communication systems.

The outline of the paper is as follows. In Sec. \ref{pre} we recall the basic notions of the connection between  Choi-Jamiolkowski isomorphism and Kraus decomposition of a quantum channel for finite dimensional systems.
In Sec. \ref{tech} we outline  our general technique to develop  operator-sum
representations for continuous-variable systems using the one-sided channel action on two-mode squeezed vacuum states. In Sec. \ref{kfock} we derive a Kraus representation for quantum-limited bosonic Gaussian channels
in the Fock basis. We focus on the quantum-limited attenuator channel and deal with the rest of the channels in a similar way in the appendices.  
In Sec. \ref{kcoh} we obtain rank-one Kraus operators for all entanglement-breaking Gaussian channels
and prove that at its respective entanglement breaking threshold, these operators are continuous-indexed and unique. 
We conclude in Section \ref{con}. 


\section{Choi-Jamiolkowski isomorphism and Kraus operators \label{pre}}
We recall some basic facts of finite-dimensional channels. Let 
$\Phi$ be a channel that transforms
density matrices of system $A$ with Hilbert space $\mathcal{H}_A$ to
density matrices of system $B$ with Hilbert space $\mathcal{H}_B$.
Positivity and trace-preserving properties of $\Phi$  are encoded as $ \Phi(\hat{\rho}^A)  = \hat{\rho}^B \geq 0, ~\forall~ \hat{\rho}^A \geq 0$ and ${\rm Tr}\,\hat{\rho}^B = {\rm Tr}\,\hat{\rho}^A$, respectively.
Additionally, due to complete positivity 
\begin{align}
\hat{\rho}^{BR} = \left(\Phi \otimes {1\!\!1}_{R}\right)(\hat{\rho}^{AR}) \geq  0,
\label{ir1}
\end{align}
where $\hat{\rho}^{AR}$ is any density operator acting on the enlarged Hilbert space $\mathcal{H}_A\otimes\mathcal{H}_R$, and ${1\!\!1}_{R}$ is the identity map on an arbitrary auxiliary (reservoir) system $R$.
In the special case where the output state  $\hat{\rho}^{BR}$ is separable {\em for every} input state $\hat{\rho}^{AR}$, $\Phi$ is called an entanglement-breaking  channel\,\cite{horodecki03}. From now on we assume that ${\cal H}_B = {\cal H}_A$ for simplicity, and so we use only label $A$ to denote the system.

The fact that any channel $\Phi$ necessarily satisfies Eq.\,(\ref{ir1}), or equivalently is realized
as a unitary dilation (see  Appendix \ref{unirep}),
implies that it admits an {\em operator-sum or Kraus representation}\,\cite{ecg61,kraus70,choi75}
\begin{align}
\Phi(\hat{\rho}^{A})= \sum_{k} A_{k}\, \hat{\rho}^{A}\, A_{k}^{\dagger}, \,\,\,\, \sum_{k} A_{k}^{\dagger} A_{k} ={1\!\!1}_A,
\label{ir3}
\end{align}
the Kraus operators $\{A_k \}$ being independent of $\hatr{A}$.
While the first relation renders the complete positivity property manifest,
the second (resolution of unity) is equivalent to the trace preserving requirement.

The Choi-Jamiolkowski  isomorphism between channels and bipartite  states\,\cite{jamiolkowski,choi75} can be exploited to extract an operator-sum representation directly from the
complete positivity requirement\,(\ref{ir1}). Consider the bipartite operator
\begin{align}
\hat{\Gamma}^{AR} = d\,|\Psi_{\rm max}\rangle \langle
\Psi_{\rm max} | = \sum_{i,j =0}^{d} E_{ij} \otimes E_{ij},
\label{max}
\end{align}
where $\{E_{ij} = |i \rangle \langle j|\}$ is the Kronecker basis for
operators of a $d$-dimensional system $A$, and 
$|\Psi_{\rm max}\rangle =\frac{1}{\sqrt{d}}\sum_{i=1}^{d}|i\,i\rangle$ is
a maximally entangled state. Note that $\hat{\Gamma}^{AR}$ is unnormalised since ${\rm Tr} \, \hat{\Gamma}^{AR} = d$, and this fact is very crucial in our analysis in subsequent sections. The local action of $\Phi$ on subsystem $A$ of $\hat{\Gamma}^{AR}$ gives the unnormalized operator 
\begin{align}
\hat{\rho}^{AR} &= (\Phi \otimes {1\!\!1} )(\hat{\Gamma}^{AR}) =  \sum_{i,j}
\Phi(E_{ij}) \otimes E_{ij},
\label{ir5}
\end{align}
and the Choi-Jamiolkowski state corresponding to channel $\Phi$ is given by 
\begin{align}
\hat{\sigma}^{AR} &:=  \frac{1}{d}\hat{\rho}^{AR}.  
\label{ir5b}
\end{align}

Now positivity of $\hat{\rho}^{AR}$ guarantees the convex rank-one 
 decomposition
\begin{align}
\hat{\rho}^{AR} = \sum_k |A_k \rangle
\langle A_k|.
\label{ir5a}
\end{align}
The bipartite {\em vectors} $|A_k\rangle =  \sum_{\ell,m}
a^{(k)}_{\ell,m} |\ell\rangle \otimes | m\rangle$ are unnormalized, and its projections are written as
$|A_k \rangle \langle A_k| = [A_k \otimes {1\!\!1}]\,
\hat{\Gamma}^{AR}\, [A_k^{\dagger} \otimes {1\!\!1}]$,
where {\em the operators} $\{A_k\}$ are $A_k = \sum_{\ell,m} a^{(k)}_{\ell,m} |\ell \rangle \langle m|$.
That is, the isomorphism \cite{jamiolkowski} between bipartite vectors $|A_k \rangle \in
{\cal H}_{A} \otimes {\cal H}_{R}$ and linear (single party) operators
$A_{k}\,:\,\,{\cal H}_{A} \to {\cal H}_{R}$
enables one to construct
{\em the Kraus operator $A_k$ from the bipartite vector
$|A_k\rangle$ by simply flipping the second ket to a bra}. Note that working with the unnormalised operator 
$\hat{\Gamma}^{AR}$ instead of the maximally entangled state in Eq. (\ref{max})
contributes to this simple correspondence. 

It is clear from Eq.\,(\ref{ir5a})
that for every pure state decomposition of
$\hat{\rho}^{AR}$ we obtain a corresponding operator-sum representation of the channel $\Phi$. Additionally, any decomposition of $\hat{\sigma}^{AR}$ ($\hat{\rho}^{AR}$) can be understood as resulting from a certain measurement on its purified state as detailed in Appendix \ref{cohrep}. Finally, note that the separability of $\hat{\sigma}^{AR}$
implies that $\Phi$ is entanglement breaking and that it has a rank-one operator-sum representation \cite{horodecki03}.

\section{Squeezed state approach  to Kraus decomposition \label{tech}}
Inspired by the Choi-Jamiolkowski approach, we develop our method for obtaining Kraus operators of continuous-variable channels.  We first develop the analogous steps for obtaining Kraus operators of continuous-variable channels that is inspired by the finite dimensional case outlined in Sec. \ref{pre}. The two-mode finitely-squeezed states play an important role here since they have an important property that they have full schmidt-rank and tend to the maximally entangled state in the limit of arbitrarily large squeezing. Consider the unnormalised operator
\begin{align}
\Gamma_r := {\rm sech}^{-2}\,r \, \proj{\Psi_r} = \sum_{m,n} ({\rm tanh}r)^{m+n} E_{mn} \otimes E_{mn}, 
  \end{align}
that is analogous to $\Gamma^{AR}$, with suitable weights. Next we consider the one-sided channel action on $\Gamma_r$ and we obtain the unnormalized operator
\begin{align}
\hr{}(r) := (\Phi_c \otimes {1\!\!1}) [\Gamma_r],
\label{oso1}
\end{align}
and 
\begin{align}
  \hat{\sigma}(r) := (\Phi_c \otimes {1\!\!1})\,\proj{\Psi_{r}}
\label{3.2b}
\end{align}
 plays the role of the Choi-Jamiolkowski state $\hat{\sigma}^{AR}$.

Comparing Eq. (\ref{oso1}) with Eq. (\ref{ir5}), it is evident that
we need to evaluate a pure state decomposition of $\hat{\rho}(r)$ to obtain the equivalent of Eq. \eqref{ir5a} and then take the
limit $r \to \infty$. Note that while in the finite dimensional case $\hr{}^{AR}$ and $\hat{\sigma}^{AR}$ are connected by a simple scalar factor `d', in the continuous-variable case one has the factor $\rm{sech}^2{r}$ that connects $\hr{}(r)$ and $\hat{\sigma}(r)$, and an important subtlety that one has to also take the limit of arbitrary large squeezing. 


Our main examples for demonstration of the method are single-mode bosonic Gaussian channels. It turns out that any Gaussian channel $\Phi$ can be decomposed into an initial Gaussian unitary $U_1$ followed by a \textit{canonical channel} $\Phi_c$ defined by a
canonical matrix pair $(X,\, Y)$, and a final Gaussian unitary $U_2$, i.e., $\Phi=U_2\circ\Phi_c\circ U_1$. It therefore suffices to obtain Kraus operators $\{T[{\Phi_c}]\}$ of $\Phi_c$ where $T[\Phi] = U_2\,T[{\Phi_c}]\,U_1$ are the Kraus operators of $\Phi$. We refer the reader to Appendix \ref{gausstuff} for further details on Gaussian states and channels.

\begin{figure}
\begin{center}
\includegraphics[width=\columnwidth]{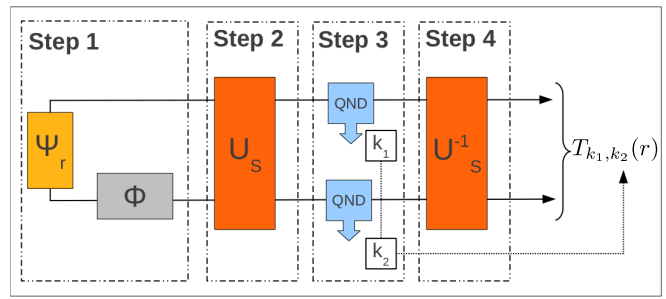}
\end{center}
\caption{The Choi-Jamiolkowski inspired technique to derive the Kraus operators
of a channel is divided in four steps: (1) send one mode of a two-mode squeezed vacuum state
$\ket{\Psi _{r}}$ through channel $\Phi$, (2) apply the Gaussian unitary operation $U_S$
that brings $V(r)$ (the covariance matrix of the total output state) to its diagonal canonical form, (3) apply a quantum-non-demolition  measurement to each mode, which output a set of classical data $(k_1,k_2)$, giving the indices of the Kraus operators and a quantum state $\ket{k_1}\otimes\ket{k_2}$, and (4) undo the unitary $U_S$ giving the final
set of vectors $\ket{{T}_{k_1,k_2}(r,\Phi)}$ from which one obtains the operators ${T}_{k_1,k_2}(r,\Phi)$. Then in the limit of large squeezing we obtain the final Kraus operators ${T}_{k_1,k_2}(\Phi) = \lim_{r \to \infty} {T}_{k_1,k_2}(r,\Phi)$ for the channel $\Phi$. }
\label{fig2}
\end{figure}

To develop the required decomposition of $\hat{\sigma}(r)$ (and simultaneously also for $\hr{}(r)$)
is divided into four elementary steps as depicted in Fig.\,\ref{fig2}.
Since all the states and channels that are involved in the procedure are Gaussian, we can directly work at the level of covariance matrices. The first step is to calculate the covariance matrix $V(r)$
of $\hat{\sigma}(r)$. 
Now $\hat{\sigma}(r)$ results from sending one mode of the (input) two-mode
squeezed vacuum state $|\Psi_r\rangle$ with variance matrix
\begin{align}
V_{\text{sq}}(r) = \left(
\begin{matrix}
\cosh{2r}\, {1\!\!1}_2 & \sinh{2r}\, \sigma_3 \\
\sinh{2r}\, \sigma_3  & \cosh{2r}\, {1\!\!1}_2
\end{matrix}
\right)
\end{align}
through the channel $\Phi_c$ (specified by $(X,Y)$), and its covariance matrix reads as  
\begin{align}
V(r) = \left(
\begin{matrix}
\cosh{2r} \, X^TX +Y && \sinh{2r} \, X^T \sigma_3 \\
\sinh{2r} \, \sigma_3 X && \cosh{2r}\, {1\!\!1}_2
\end{matrix}
\right).
\end{align}
Note that requiring $V(r)$  respect the uncertainty principle $V + i \Omega \geq 0$ \cite{dutta94a} for all $r$
is equivalent to the complete-positivity condition of $\Phi_c$. Channels which saturate this condition are known as quantum-limited channels. 
Further, the separability of the state corresponding to $V(r)$, given by the condition $V(r)^{PT} + i\Omega \geq 0$  (PT stands for partial transpose \cite{simon00})
for arbitrary $r$ is equivalent to the condition that the channel $\Phi_c$ is entanglement breaking. Channels which saturate this condition are said to be at its entanglement-breaking threshold. \\

\noindent
{\bf Remark}: The reason we are able to obtain properties of the channel through the state $\hat{\sigma}(r)$ corresponding to $V(r)$ is because taking the limit of arbitrary large squeezing of $\ket{\Psi_r}$ one obtains  the `maximally entangled state'. For Gaussian channels, it is an interesting observation that both the complete-positivity and entanglement-breaking condition are independent of the value of $r$, and are therefore obtained even for finitely squeezed test states.  \hfill $\blacksquare$\\
 
The second step is to apply the appropriate two-mode Gaussian unitary  operator $U_S$
to $\hat{\sigma}(r)$ that brings $V(r)$ to its diagonal canonical form\,\cite{dutta94a,dutta94b}
\begin{align}\label{k8}
V_{\text{can}}(r) \equiv S\,V(r)\, S^T =
\text{diag}(\nu^+, \nu^+, \nu^-, \nu^-),
\end{align}
where $\nu^+$ and $\nu^-$ are the thermal parameters and also the symplectic eigenvalues of $V(r)$. Note that the superscripts $+,-$ on the thermal parameters $\nu^{+}, \nu^-$ denote the first and second mode respectively.
We now have $\hat{\sigma}_{\rm can}(r)=
U_S\,\hat{\sigma}(r)\,U_S^{\dagger}$ to be a product of thermal states
\begin{align}\label{k9}
\hat{\sigma}_{\rm can}(r) = \hat{\sigma}_{\text{th}}(\nu^{+})\otimes\hat{\sigma}_{\text{th}}(\nu^{-}).
\end{align}

The third step of our procedure is to obtain a
decomposition of this product thermal state either in
the Fock basis or in the  coherent state basis depending on the context.
For instance, Eq. (\ref{k9}) is diagonal in the
Fock basis $\{ | k_1 \rangle \otimes | k_2 \rangle \}$ and, as we will see, the spectral
decomposition of the thermal states (eventually) determines a set of Kraus operators.
One could alternatively see this stage as two quantum-non-demolition  measurements,
one on each mode, that output a set of classical data $(k_1,k_2)$, giving the indices
of the Kraus operators  corresponding to the Fock state $\ket{k_1}\otimes\ket{k_2}$.

The fourth and final step of the procedure is to undo the unitary $U_S$ to recover a
decomposition of $\hat{\sigma}(r)$ itself\,:
\begin{align}
\hat{\sigma}(r) &= U_{S}^{\dagger} \, \hat{\sigma}_{\rm can}(r) \, U_S = U_{S}^{\dagger}\, \left( \hat{\sigma}_{\text{th}}(\nu^{+})\otimes
\hat{\sigma}_{\text{th}}(\nu^{-})\right)\, U_{S} \nonumber\\
& = \sum_{k_1,k_2} \bra{k_1}\hat{\sigma}_{\text{th}}(\nu^{+}) \ket{k_1} ~~\bra{k_2}\hat{\sigma}_{\text{th}}(\nu^{+}) \ket{k_2} \nonumber\\
& \hspace{1cm} \times U_S^{\dagger} \proj{k_1,k_2} U_S \label{eq3.6}\\
&= {\rm sech}^2\,r\,( \hat{\rho}(r)), \text{ where } \nonumber 
\end{align}
\begin{align}
 \hat{\rho}(r) &= \sum_{k_1, k_2}\proj{{T}_{k_1,k_2}(r,\Phi)}, \text{ with } \nonumber \\
 &\ket{{T}_{k_1,k_2}(r,\Phi)} \,\propto\, U_S^{\dagger}\,\ket{k_1}\otimes\ket{k_2}.
\end{align}
In Eq. \eqref{eq3.6} we used the fact that the thermal state is diagonal in the Fock basis.      
The Kraus operators $\{{T}_{k_1,k_2}(r,\Phi)\}$ of the corresponding trace-non-preserving completely positive map are read out from the respective bipartite vectors $\{\ket{{T}_{k_1,k_2}(r,\Phi)} \}$
by flipping the second ket to a bra, an instance of the Choi-Jamiolkowski isomorphism. Finally, taking the limit $r \to \infty$ we obtain
the required Kraus operators of the Gaussian channel under consideration.\\

\noindent
{\bf Remark.}
It should be appreciated that steps 2 to 4 constitute a convenient device for obtaining a pure state decomposition of the bipartite state $\hat{\sigma}(r)$ (and hence, also of $\hat{\rho}(r)$). \hfill $\blacksquare$

\section{Kraus decomposition in the Fock basis\label{kfock}}
We begin our implementation of the above procedure for the derivation of Kraus representation of attenuators and amplifiers in the Fock basis. We focus on the quantum-limited attenuator channel and derive its Kraus operators in full detail. We then provide a Kraus decomposition for the quantum-limited amplifier in Appendix \ref{qlamp} and for the quantum-limited phase-conjugation channels in \ref{qlpc}. It turns out that we recover the same Kraus operators for these quantum-limited channels as derived earlier in Ref. \cite{gaussch}. This technique is then extended to noisy amplifier, attenuator, and phase conjugation channels, in Appendix \ref{noisyc}.

\subsection{Attenuator and amplifier channels}
The attenuator channel $\mathcal{C}_1(\kappa,\alpha)$  and the amplifier channel $\mathcal{C}_2(\kappa,\alpha)$,
with losses/gain parameter $\kappa$ and noise parameter $\alpha$, are  specified by  matrices
$(X,Y) =(\kappa 1\!\!1, \alpha 1\!\!1)$, with $\kappa \leq 1$ and $\kappa \geq 1$
respectively, and $\alpha \geq |\kappa^2-1|$. Under the one-sided action of the channel
on $|\Psi_r\rangle$ the variance matrix of the output Gaussian state
of step $1$ of our procedure in Sec. \ref{tech} is given by
\begin{align}\label{b1}
{V}^{b/a}(r,\kappa,\alpha) = \left[
\begin{matrix}
(\kappa^2\cosh{2r}+\alpha)1\!\!1_2  & \kappa \sinh{2r} \,\sigma_3 \\
\kappa \sinh{2r}\, \sigma_3 & \cosh{2r}\,1\!\!1_2
\end{matrix}
\right].
\end{align}
Here the superscript $b/a$ denotes that
we are dealing  with either the beam splitter (attenuator) or the amplifier channel.
The pair $(\kappa1\!\!1, \alpha1\!\!1)$
represents a Gaussian channel  only if
$V^{b/a}(r,\kappa,\alpha)$ obeys the uncertainty principle
for arbitrary $r$, i.e.,  both its symplectic eigenvalues $\nu^{+}_{b/a}$ and $\nu^{-}_{b/a}$ are
$\geq 1$; but this is seen to hold if and only if $\alpha \geq |\kappa^2-1|$, {\em independent of $r$}.
To determine for what values of $\alpha$ will the channel be entanglement breaking,
we apply the partial transpose test on ${V}^{b/a}(r, \kappa,\alpha)$ of Eq.\,\eqref{b1}
and see if it corresponds to a separable state.
As shown in\,\cite{simon00}, this is the case when
the symplectic eigenvalues $\tilde{\nu}^{\pm}_{b/a}$ of the partially transposed
variance matrix $V^{b/a}(r,\kappa,\alpha)^{PT}$ are $\geq 1$.
It is readily verified that this happens if and only if $\alpha \geq \kappa^2 +1$ \cite{holevo08}, {\em independent of $r$} as well.

Note that the diagonal blocks of ${V}^{b/a}(r,\kappa,\alpha)$ 
are multiples of the unit matrix, while its off-diagonal block is
proportional to $\sigma_3$. This readily suggests that it can be  diagonalized in step $2$ by a
two-mode squeezing transformation $S(\mu) \in {\rm Sp}(4, {\mathbb R})$ where
\begin{align}\label{b2a}
 S(\mu)=\left[ \begin{array}{cc}
\cosh{\mu}\, 1\!\!1_2 & -\sinh{\mu}\, \sigma_3\\
-\sinh{\mu}\, \sigma_3 &\cosh{\mu} \,1\!\!1_2
\end{array}\right],\,\,\, \\
\text{tanh}\,{2\mu}= \frac{2\kappa \sinh{2r}}{\cosh{2r} + \kappa^2 \cosh{2r} +\alpha}.
\label{b2b}
\end{align}
The squeezing parameter $\mu$ of the transformation is thus
a function of $r$, $\kappa$, and $\alpha$\,, i.e.,
$\mu\equiv \mu(r,\kappa,\alpha)$.
The (doubly degenerate) symplectic eigenvalues of ${V}^{b/a}(r,\kappa,\alpha)$  are
\begin{align}\label{b4}
\nu^{\pm}_{b/a}&=\frac{1}{2} \left[ z_{b/a}
\pm w_{b/a} \right], \nonumber \\
z_{b/a}&=[(\alpha+(\kappa^2+1) \cosh{2r})^2-4\kappa^2\sinh^2{2r}]^{1/2}, \nonumber \\
w_{b/a} &= \alpha+(\kappa^2-1) \cosh{2r}.
\end{align}

While we treated the beam splitter and amplifier channels together in steps 1 and 2 it is convenient to treat them separately for the remaining steps 3 and 4. We first begin with the case of the quantum-limited attenuator channel for ease of presentation.

\subsubsection{Quantum-limited attenuator ${\cal C}_1(\kappa)$}
Quantum-limited channels are those for which
$\alpha$ assumes the least value for a given $X$ as dictated by the complete positivity requirement (see also Appendix \ref{gausstuff}).
For the quantum-limited attenuator channel ($\kappa \leq 1$) we have
$\alpha=1-\kappa^2$, so that
$\nu^{+}_{b}=1$ and $\nu^{-}_{b} =(1-\kappa^2) \cosh{2r}+\kappa^2$,
and Eq.\,\eqref{b2b} reduces to
${\rm tanh}\,\mu = \kappa\, {\rm tanh}\,r$. 
The symplectic transformation of Eq.\,\eqref{b2a} of step 2  is given by
\begin{align}
&S[\mu(r, \kappa)] = \nonumber\\
&\frac{1}{\sqrt{1-\kappa^2 \tanh^2{r}}}\left[
\begin{array}{cc}
1\!\!1_2 & -\kappa \tanh{r}\, \sigma_3\\
-\kappa \tanh{r}\, \sigma_3 & 1\!\!1_2
\end{array}
\right].
\label{b5}
\end{align}
Thus in step 3  we find
$\hat{\sigma}^{b}_{\rm can}(r, \kappa)$ is a tensor product of the ground state in the first mode and a thermal
state with parameter $\nu_{b}^{-}$ in the second mode. 

We decompose $\hat{\sigma}_{\rm can}^{b}(r, \kappa)$ in the Fock basis as
\begin{align}
\hat{\sigma}_{\rm can}^{b}(r,\kappa) &= {\rm sech}^2 r\,N_{b}(r,\kappa) \sum_{n=0}^{\infty} \left( x(\nu^{-}_{b})\right)^{2n}
|0,n \rangle\langle 0, n|,\,\, \nonumber \\
N_{b}(r, \kappa)&=[1-\kappa^2\tanh^2{r}]^{-1},\,\, \nonumber\\
x({\nu}^{-}_{b})  &\equiv  \left[\frac{\nu^{-}_{b} -1}{\nu^{-}_{b} +1}\right]^{1/2} = \left[\frac{1-\kappa^2}{\coth^2{r}-\kappa^2}\right]^{1/2}.
\label{b6}
\end{align}
Note that we have already factored out ${\rm sech}^2 r$.
In the final step 4 we undo the unitary two-mode squeezing of Eq.\,\eqref{b5} to obtain
\begin{align}\label{b7}
\hat{\sigma}^{b}(r,\kappa) &= U_S^{\dagger}\,\hat{\sigma}^{b}_{\rm can}(r,\kappa) \,U_S = ({\rm sech}^2 r)\, \hat{\rho}^b(r,\kappa), \nonumber\\
\hat{\rho}^b(r,\kappa)&= \sum_{n=0}^{\infty}\proj{{T}_{n}^{b}(r,\kappa)},
\end{align}
where
\begin{align}
|{T}_{n}^{b}(r,\kappa)\rangle &= \sqrt{N_{b}(r, \kappa)} \,
\left(x(\nu^{-}_{b})\right)^n\, U_S^{\dagger}|0,n\rangle \nonumber\\
 &=\sqrt{N_{b}(r, \kappa)} \,
\left(x(\nu^{-}_{b})\right)^n\, \nonumber\\
&~~~~\times \sum_{m=0}^{\infty} \sqrt{\binom{m+n}{n}}\,
\frac{\tanh^{m}{\mu}}{\cosh^{n+1}{\mu}}|m,m+n \rangle.
\label{b7a}
\end{align}
In Eq. \eqref{b7a} we used the matrix elements of the two-mode squeezing operator in the Fock basis (see Eq. (5.4) of Ref. \cite{gaussch}).
We associate Kraus operators ${T}^b_n(r,\kappa)$ with the unnormalized bipartite states
$|{T}_{n}^{b}(r,\kappa)\rangle$ by flipping the second ket to a bra, i.e.,
\begin{align}
\label{b8}
{T}_{n}^{b}(r,\kappa) &= \sqrt{N_{b}(r,\kappa)}\,\left(x(\nu^{-}_{b})\right)^n \sum_{m=0}^{\infty} \sqrt{\binom{m+n}{n}}\, \nonumber \\
&  ~~~\times
\frac{\tanh^{m}{\mu}}{{\rm cosh}^{n+1}{\mu}} |m\rangle \langle m+n|, ~~n=0,1,2,\cdots.
\end{align}
The resulting  completely positive map ${\cal C}_1(r,\kappa)$, with Kraus
operators $\{{T}_{n}^{b}(r, \kappa)\}$,  is not yet
trace-preserving as we are still in the finite squeezing domain. This fact is made transparent by evaluating the operator
$\sum_{n} T_n^b(r,\kappa)^{\dagger}T_n^b(r,\kappa)$, which should be a resolution of identity for a channel. We have
\begin{align}
&T_n^b(r,\kappa)^{\dagger}T_n^b(r,\kappa) \nonumber\\
&=\frac{N_b(r,\kappa)}{\cosh^2{\mu}} \sum_{m=0}^{\infty} \left(\frac{x(\nu_b^{-})}{\cosh{\mu}}\right)^{2n} \,\binom{m+n}{n}\nonumber\\
&~~~~\times  \,(\tanh{\mu})^{2m} |m+n \rangle  \langle m+n|,
\end{align}
which is diagonal in the Fock basis for all $n$, $r$.
It is easy to see that for a fixed $m+n=t$, the coefficient of $|t \rangle \langle t |$ in the
sum $\sum_{n=0}^{\infty} T_n^b(r,\kappa)^{\dagger}T_n^b(r,\kappa)$
is given by
\begin{align}
\frac{N_b(r,\kappa)}{\cosh^2{\mu}}
\sum_{n=0}^{t} \left(\frac{x(\nu_b^{-})}{\cosh{\mu}}\right)^{2n} \,\binom{t}{n}
(\tanh{\mu})^{2(t-n)},
\end{align}
which leads to
\begin{align}
&\sum_{n=0}^{\infty} T_n^b(r,\kappa)^{\dagger}T_n^b(r,\kappa)\nonumber\\
&=\frac{N_b(r,\kappa)}{\cosh^2{\mu}} \sum_{t=0}^{\infty} \left[\frac{(x(\nu_b^{-}))^2}{\cosh^2{\mu}} + \tanh^2{\mu}\right]^t \,|t\rangle \langle t|.
\label{4.13}
\end{align}
Substituting the expressions for $N_b(r,\kappa),\,x(\nu_b^{-})$ from Eq.\,\eqref{b6} and for  $\tanh{\mu},\,\cosh{\mu}$ from Eqs.\,\eqref{b2a} and \eqref{b5}, we have the Fock matrix elements of the operator in Eq. \eqref{4.13} to be
\begin{align}
\left(\sum_{n=0}^{\infty} T^{b}_n(r,\kappa)^{\dagger} T^b_n(r,\kappa)\right)_{j \ell} = \delta_{j\ell} \,(\tanh^2{r})^j.
\label{btp}
\end{align}

\noindent
{\bf Remark.} While this expression is not a resolution of identity for any  $r$, the diagonal entries assume the {\em universal form} of the spectrum of a thermal state. Indeed, one could have anticipated this expression even without the need for the preceding computation, since one knows in advance that it ought to coincide with the reduced state of the mode on which the identity channel acts in the Choi-Jamiolkowski state in Eq.\,\eqref{3.2b}. Our computation demonstrates consistency. \hfill $\blacksquare$

Now going back to $T^{b}_{n}(r, \kappa)$ in Eq.\,\eqref{b8},
in the limit $r \to
\infty$, we have $x(\nu^{-}_{b}) \to 1$ and $N_{b}(r,\kappa) \to ({1-\kappa^2})^{-1}$.
Further,
${\rm  tanh}[2\,\mu(r,\kappa)] \rightarrow {\rm  tanh}[2\,\mu_0(\kappa)] ={2\kappa}/({1+\kappa^2})$,
${\rm sech}[\mu(r,\kappa)]\rightarrow {\rm sech}[\mu_0(\kappa)] =\sqrt{1-
  \kappa^2}$, and ${\rm tanh}[\mu(r,\kappa)]\rightarrow {\rm tanh}[\mu_0(\kappa)]=\kappa$, where $\mu_0(\kappa) := \lim\limits_{r\to \infty} \mu(r,\kappa)$.
The final Kraus operators of the quantum-limited
beam splitter (attenuator) channel are thus  read off from Eq.\,\eqref{b8} as
\begin{align}\label{b9}
B_n(\kappa)&= \lim_{r \to \infty} T^b_n(r,\,\kappa)\nonumber\\
&=\sum_{m=0}^{\infty} \sqrt{\binom{m+n}{n}}\,
\kappa^{m} \left(\sqrt{1-\kappa^2 }\right)^{n} |m \rangle \langle m+n|, \nonumber \\
&\hspace{3cm}\,\,n=0,1,2,\cdots.
\end{align}
Consequently, by Eq. \eqref{btp}, we recover in the $r \to \infty$ limit the  expected trace-preserving property
\begin{align}
\lim_{r \to \infty} \,\sum_{n=0}^{\infty} T_n^b(r,\kappa)^{\dagger} T^b_n(r,\kappa) = \sum_{n=0}^{\infty} B_n(\kappa)^{\dagger} B_n(\kappa) =1\!\!1.
\end{align}

\subsubsection{Quantum limited amplifier ${\cal C}_2(\kappa)$}
We  proceed in a similar way to obtain the operators $T_n^a(r,\kappa)$  from which we recover in the limit $r \to \infty$ the Kraus operators  for the quantum-limited amplifier channel
\begin{align}
&A_n(\kappa)=\lim_{r \to \infty} T^a_n(r,\,\kappa)\nonumber\\
&=\sum_{m=0}^{\infty} \sqrt{\binom{m+n}{n}}
\left(\frac{1}{\kappa}\right)^{m+1}\!\! \left(\frac{\sqrt{\kappa^2-1 }}{\kappa}\right)^{n}
\!\!|m+n \rangle \langle m|, \nonumber \\
&\hspace{3cm}\,\,n=0,1,2,\cdots.
\end{align}
We provide the entire derivation in Appendix \ref{qlamp}.

\subsection{Phase conjugation ${\cal D}(\kappa)$}
The  phase-conjugation channel, denoted by ${\cal D}(\kappa;\alpha)$, is specified by the matrix pair $(\kappa \sigma_3;\alpha 1\!\!1)$ with $\kappa >0, \alpha \geq \kappa^2+1$. The channel is quantum-limited when $\alpha = \kappa^2 +1$ (see Appendix \ref{qlpc}). Further, the phase conjugation channel is always entanglement breaking irrespective of the value of the channel parameters. Following a similar analysis to the quantum-limited attenuator channel, one obtains for the  quantum-limited phase-conjugation channel the Kraus operators 
\begin{align}
C_{n}(\kappa) &= \lim_{r \to \infty} T^c_n(r,\,\kappa)\nonumber\\
&=\frac{1}{\sqrt{1+\kappa^2}}
\sum_{m=0}^{n}\sqrt{\binom{n}{m}} \,\left[\frac{\kappa}{\sqrt{1+\kappa^2}}\right]^{n-m}
 \nonumber \\
& \times \left[\frac{1}{\sqrt{1+\kappa^2}} \right]^m  \,
|n-m \rangle \langle m|,\, n=0,1,\cdots.
\label{pckraus}
\end{align}
We refer the reader to Appendix \ref{qlpc} for the derivation. 
%
%

\section{Uniqueness of Kraus decomposition at the entanglement-breaking threshold\label{kcoh}}

It is well known that entanglement breaking channels admit an operator-sum representation with rank-one Kraus operators\,\cite{horodecki03,holevo08}. For finite dimensional systems the operator-sum representation of entanglement-breaking
channels are composed of sets of countable rank-one Kraus operators. The existence of a unique non-countable rank-one operator-sum representation
of entanglement-breaking channels of infinite dimensional systems was shown in \cite{holevo05} where a mathematical example was constructed.
A consequence of our divergence-free approach is the following theorem. 

{\begin{theorem}
All single-mode bosonic Gaussian channels at its respective entanglement breaking thresholds admit a continuous-indexed (non-countable) set of rank-one Kraus operators that is also unique, with the exception of the ${\cal A}_1$ class of full loss channels.
\end{theorem}

\noindent
{\bf Proof of existence.}
The rank-one operator-sum decomposition of entanglement breaking Gaussian channels at its respective thresholds was derived for the phase-conjugation, and singular channels in \cite{gaussch}, and for the amplifier channel in Ref. \cite{threec}.
Table \ref{table2} summarizes the rank-one Kraus decomposition for entanglement-breaking channels at its respective entanglement breaking thresholds. Note that the final column of the table depicts the action of the entanglement-breaking channel at the level of the Glauber-Sudarshan diagonal representation. It is manifestly transparent that all these entanglement-breaking channels output states that are classical in the quantum-optical context. Channels with this property are known as nonclassicality-breaking channels \cite{ncb,nbm,pon}.  
\begin{table}
\caption{Summary of rank-one Kraus decompositions at the entanglement-breaking thresholds of single-mode bosonic Gaussian channels. $\phi(\beta;\hr{\rm out})$ denotes the Glauber-Sudarshan diagonal quasiprobability corresponding to the output state of channel with input state $\hr{\rm in}$, and $Q(\beta;\hr{\rm in})$ denotes the Husimi $Q$-function corresponding to the input state, with the phase-space variable $\beta = (\beta_x +i\beta_y)/\sqrt{2}$. The subscript $\ket{\cdot}_{\rm coh}$ and $\ket{\cdot}_{\rm pos}$ stand for the coherent and position basis respectively, and EB denotes entanglement-breaking.}
\begin{tabular}{cccc}
\hline
Channel & EB  &Kraus &$\phi(\beta;\hr{\rm out})$   \\
$\Phi$ & threshold& operators& \\
\hline
\hline
${\cal C}_1(\kappa;\alpha)$ &$\alpha = \kappa^2 +1$ & $\ket{\kappa\,\beta} \bra{\beta}$& $\kappa^{-2} Q(\kappa^{-1} \beta;\hr{\rm in})$\\
${\cal C}_2(\kappa;\alpha)$ &$\alpha = \kappa^2 +1$ & $\ket{\kappa \,\beta} \bra{\beta}$& $\kappa^{-2} Q(\kappa^{-1} \beta;\hr{\rm in})$\\
${\cal D}(\kappa;\alpha)$ &$\alpha = \kappa^2 +1$ & $\ket{\kappa \,\beta^*} \bra{\beta}$& $\kappa^{-2} Q(\kappa^{-1} \beta^*;\hr{\rm in})$\\
${\cal B}_2(\alpha)$ &$\alpha = 2$ & $\proj{\beta}$& $Q(\beta;\hr{\rm in})$\\
${\cal A}_2(\alpha)$ &$\alpha = 1$ & $\ket{{\rm Re}\,{\beta}}_{\rm coh} \bra{\beta_x}_{\rm pos}$&$ \delta(\beta_y) \bra{\beta_x} \hr{} \ket{\beta_x}$\\
${\cal A}_1(\alpha)$ & $\alpha =1$ & $\ket{0}\bra{\beta}$ & $\delta^2(\beta)$\\ 
\hline
\end{tabular}
\label{table2}
\end{table}
We now discuss how our divergence-free approach is used to recover these Kraus operators. 

\emph{Quantum-limited phase-conjugation}: 
The family of Kraus operators $\{C_n(\kappa)\}$ obtained for the quantum-limited phase conjugation channel in Eq. \eqref{pckraus} are precisely of rank $n+1$. This implies that a different family of Kraus operators exist that is rank-one. We now decompose the thermal state of Eq.\,\eqref{pc5} [of Appendix \ref{qlpc}] in step 3 in the continuous and over-complete coherent state `basis', instead of the Fock basis, to obtain
\begin{align}\label{r1}
  &\hat{\sigma}_{\text{can}}^{c}(r, \kappa) = \hat{\sigma}(\nu_c^+) \otimes \proj{0} \nonumber\\
  &= \frac{2}{\nu_c^{+}-1}\int\frac{d^2\beta}{\pi}\, \exp \left[-\frac{2
    |\beta|^2 }{\nu_c^{+}-1}\right]  \proj{\beta}\otimes\proj{0},
\end{align}
where $\nu^+_c = {\rm cosh}{2r}\, (\kappa^2+1) + \kappa^2$.

The state $\ket{\beta} \otimes \ket{0}$ is mapped to
$\ket{\cos{\theta(r,\kappa)}\,\beta} \otimes \ket{-\sin{\theta(r,\kappa)}\,\beta}$ by the symplectic diagonalization unitary $U^\dagger_S[\theta(r,\kappa)]$ in step 3, which corresponds to a two-mode beam splitter with parameter $\theta(r,\kappa)$. The symplectic transformation corresponding to this beamsplitter is provided in Eq. \eqref{pc2} of Appendix \ref{qlpc}, where $\cos{\theta} = \kappa [\kappa^2 + {\rm tanh}^2{r}]^{-1/2}$. So we obtain 
\begin{align}
\hat{\sigma}^{c}(r,\kappa) &= {\rm sech}^2 r \,\int \frac{d^2\beta}{\pi}
|{T}_{\beta}^{c}(r, \kappa) \rangle \langle {T}_{\beta}^{c} (r, \kappa)|,\,\,{\rm where}
\nonumber \\
\ket{{T}_{\beta}^{c} (r,\kappa )} &= \frac{1}{\sqrt{1+\kappa^2 - {\rm sech}^2 r}}
\exp \left[-\frac{|\beta|^2}{(\nu^{+}_{c}(r,k)-1)} \right]  \nonumber \\
& \times |\beta \cos{\theta(r, k)}  \rangle \otimes |-\beta\sin{\theta(r,k)} \rangle.
\label{r2}
\end{align}
In the limit $r \rightarrow \infty$ we obtain, by flipping the second ket of
$|{T}_{\beta}^{c}(r, \kappa) \rangle$  to a bra, the channel's rank-one Kraus operators
\begin{align}\label{r3}
T_{\beta}^{c}(\kappa)=\frac{1}{\sqrt{(1+\kappa^2)}}| (\kappa/\sqrt{1+\kappa^2})\,\beta \rangle\langle
  (1/\sqrt{1+\kappa^2})\,\beta^*|.
\end{align}
For simplicity, we perform a change of variables  $\beta/\sqrt{1 + \kappa^2}$ to $\beta^*$, and we obtain
\begin{align}\label{r3b}
T_{\beta}^{c}(\kappa)= \ket{\kappa\beta^*} \bra{\beta},
\end{align}
where the trace preserving condition reads
\begin{align}
\int \frac{d^2\beta}{\pi}\, T_{\beta}^{c\,\dagger}(\kappa) \,T_{\beta}^{c}(\kappa) = {1\!\!1},
\end{align}
and the Kraus decomposition is written as
\begin{align}
\Phi(\hat{\rho})= \int \frac{d^2\beta}{\pi}\, T_{\beta}^{c}(\kappa)\, \hat{\rho}\, T_{\beta}^{c\,\dagger}(\kappa).
\label{r4}
\end{align}
So a quantum-limited phase-conjugation channel
is strictly equivalent to a scheme where the input to the channel is measured in the coherent state basis,
and for outcome ${\beta}$ the coherent state
$\ket{\kappa{\beta}^*}$ is prepared.



\emph{Beam splitter and amplifier channels }:  As seen before, the beam splitter and amplifier channels are entanglement breaking for
$\alpha \geq \kappa^2+ 1$. 
It is easy to see, by comparing the variance matrices in Eqs.\,\eqref{b1} and\,\eqref{pc1}
that $\hat{\sigma}^{b/a}(r,\kappa,\alpha)$
of an attenuator or amplification channel with $\alpha\geq\kappa^2+1$
is the partial transpose of $\hat{\sigma}^{c}(r, \kappa, \alpha)$ which corresponds to a phase-conjugation channel with the same channel parameters $\kappa,\alpha$.
Therefore, to obtain the Kraus representation of an attenuator
or amplification channel at the entanglement breaking threshold, we simply apply
a partial transpose to the expression
in Eq.\,\eqref{r2}, leading to a rank-one Kraus representation given by
\begin{align}
T_{\beta}^{b/a}(\kappa;1+\kappa^2)=\ket{\kappa \beta} \bra{\beta}.
\label{5.7}
\end{align}
Here $T_{\beta}^{b/a}(\kappa;\alpha)$ corresponds to the Kraus operators of a noisy amplifier or beam splitter channel. Note that the $r \to \infty$ limit has already been applied and the transposition was performed in the Fock basis.  
So an amplifier or attenuator channel at the entanglement breaking threshold
is strictly equivalent to a scheme
where the input to the channel is measured in the coherent state basis,
and for outcome ${\beta}$ we prepare $\ket{\kappa{\beta}}$. 

\emph{${\cal A}_1(\alpha)$, ${\cal A}_2(\alpha)$, ${\cal B}_1(\alpha)$, and ${\cal B}_2(\alpha)$ channels}: 
We recall that the transmission and noise matrix pair $(X,Y)$  are $(0,\alpha 1\!\!1)$ for ${\cal A}_1(\alpha)$, $((1\!\!1 + \sigma_3)/2, \alpha 1\!\!1)$ for ${\cal A}_2(\alpha)$, $(1\!\!1, \alpha (1\!\!1 + \sigma_3)/2)$ for ${\cal B}_1(\alpha)$, and $(1\!\!1, \alpha 1\!\!1)$ for ${\cal B}_2(\alpha)$. Further, the classical noise channel ${\cal B}_2(\alpha)$ and the class ${\cal A}_1(\alpha)$ are limiting cases of the noisy beam splitter channel corresponding to
$\kappa \to 1$ (also $\kappa \to 1$ of the noisy amplifier) and  $\kappa \to 0$ (also $\kappa \to 0$ of the noisy phase conjugator), respectively.

${\cal B}_2(\alpha)$ is entanglement breaking for $\alpha \geq 2$ (from Eq. \eqref{b1}).  Therefore, its respective rank-one decomposition in the coherent state basis
can be derived from the previous result of the amplifier and attenuator channels taking the corresponding limit and we obtain its Kraus operators  $\{T_{\beta} = \proj{\beta}\}$. The channel ${\cal A}_1(\alpha)$ is entanglement breaking for all $\kappa \geq 1$. So, taking the limit $\kappa \to 0$ in Eq. \eqref{5.7} we get a set of Kraus operators for the channel ${\cal A}_1(1)$ as $\{T^0_{\beta} = \ket{0} \bra{\beta}\}$. 
We consider the case of singular channels ${\cal A}_2$ 
in Appendix \ref{secsing} (and also the class ${\cal B}_1(\alpha)$), thereby completing the demonstration of proof of existence of our method for all single-mode Gaussian channels. Note that the class ${\cal B}_1(\alpha)$ are never entanglement-breaking. \hfill$\square$ 

We introduce the following lemma that is central to the uniqueness proof.

\begin{lemma}[\cite{aharanov}]
A pure state $|\psi \rangle$, when coupled to
the ground state $|0\rangle$ of an auxiliary mode and passed through a beam splitter (except full reflectivity),
is mapped to a product state if and only if $|\psi \rangle$ is a coherent state.
\end{lemma}

\noindent
{\bf Proof of Uniqueness.}
\emph{Phase-conjugation channels}: 
The rank-one Kraus operators in Eq. \eqref{r3} 
of the phase conjugation channel  were obtained as a result of
sending the thermal state coupled to the ground state through the beam splitter $U_S^{\dagger}[\theta]$. 
Any rank-one decomposition of the phase-conjugation channel
should have a corresponding separable state
\begin{equation}
\hat{\sigma}^c(r,\kappa)=\sumint p_i \ket{\psi_i}\bra{\psi_i}\otimes\ket{\phi_i}\bra{\phi_i}
\label{equation:separable}
\end{equation}
that is mapped into $\hat{\sigma}^{c}_{\rm can}(r, \kappa) = \hat{\sigma}_{\rm th}(\nu_c^+) \otimes\ket{0}\bra{0}$
after applying the unitary $U_S[\theta]$. Therefore, for any
decomposition in Eq. \eqref{equation:separable} there must be a corresponding decomposition
of $\hat{\sigma}_{\rm th}(\nu_c^+)$.
%
Then, by Lemma 1 it is clear that had we resolved the thermal mode
of $\hat{\sigma}_{\rm can}^{c}(r,\kappa)$
in any basis other than the coherent state basis,
we would end up with a decomposition of $\hat{\sigma}^{c}(r,\kappa)$
where  the constituent states {\em would not be} of
the product form, even though the state itself is separable,
which proves the uniqueness of the coherent state decomposition.

\emph{Remaining channels}: 
We have already seen that the Jamiolkowski states of the
beam splitter and amplifier channels, at its respective
entanglement breaking thresholds, are in one-to-one correspondence to the separable states
$\hat{\sigma}^{c}(r, \kappa)$ via the partial transposition operation, i.e.,
\begin{equation}
\hat{\sigma}^{b/a}(r,\kappa)= \hat{\sigma}^{c}(r,\kappa)^{PT} = \sumint p_i \ket{\psi_i}\bra{\psi_i}\otimes\ket{\phi_i}\bra{\phi_i}^{T}.
\label{equation:separable2}
\end{equation}
Since transposition maps coherent states $\ket{\alpha}$ to coherent states $\ket{\alpha^*}$,
and as explained before, any decomposition in Eq. \eqref{equation:separable2} is in one-to-one correspondence to a decomposition
of $\hat{\sigma}_{\rm th}(\nu_c^+) \otimes\ket{0}\bra{0}$. 
Lemma 1 implies again the uniqueness of the coherent state rank-one decomposition.
%
Similar arguments also hold true for the singular channel ${\cal A}_2(\alpha)$ dealt in Appendix \ref{secsing}.

{\it Pathological case}: There is however one exception in ${\cal A}_{1}(1)$.
In this case, in step 3 of the procedure, $\hat{\sigma}^{b/a}(r,1)$ itself is a product of the
ground state and a thermal state. Thus every pure state decomposition $\{\ket{\psi_i}\bra{\psi_i}\}$ of
this thermal state results in a rank-one set of Kraus operators $\ket{0}\bra{\psi_i}$,
which leads to a pathological case where any countable or non-countable basis of the input Hilbert space will lead to a
valid rank-one operator-sum decomposition of the channel.  \hfill $\blacksquare$
%

Finally, we remark that a similar procedure can be followed to obtain the rank-one Kraus decomposition for entanglement breaking channels above its threshold noise value. It turns out that the measurement-preparation description of these channels is a coherent state measurement followed by a preparation of a suitable (displaced) thermal state.

\section{conclusions \label{con}}
Our main contribution is a divergence-free  approach for obtaining Kraus decompositions of
continuous-variable channels that was motivated by the Choi-Jamiolkowski
isomorphism. The method makes effective
use of phase space techniques and two-mode finitely squeezed vacuum states, thereby avoiding technical difficulties or divergences naturally occurring in continuous-variable systems, mainly due to the non-existence of a maximally entangled state. Our method 
begins with the application of  the channel to one subsystem of a two-mode squeezed state. The pure state decompositions of the resulting state gave us the corresponding Kraus operators of the channel in the limit of arbitrarily large squeezing.  We obtained the Kraus operators of all single-mode bosonic Gaussian channels in either the Fock or coherent state basis depending on context. Also a novelty of the method was that it treated both quantum-limited and noisy Gaussian channels on an equal footing. This gave rise to new Kraus operators for noisy channels which previously required composition of Kraus operators of suitable quantum-limited channels. 

For entanglement breaking bosonic Gaussian channels we obtained a set of rank-one Kraus operators using the overcomplete basis of coherent states. 
Additionally, we demonstrated that at the corresponding entanglement breaking thresholds, these rank-one operators are continuous-indexed and unique, thereby providing a natural instance of a channel with no countable rank-one Kraus decomposition.  


The method could have practical applications to simulation of channels  and its experimental implications \cite{sim1,sim2,sim3}. 
Our method in principle is universal and hence could also prove useful in the study of non-Gaussian quantum channels of continuous-variable systems \cite{pagc,ng1,ng2,ng3}. 

\begin{acknowledgments}
\noindent R.G.-P.  is a Research  Associate  of  the  Fonds  de  la  Recherche  Scientifique  (F.R.S.-FNRS). 
\end{acknowledgments}

\appendix

\section{Unitary dilation and Kraus decompositions \label{unirep}}
Any channel $\Phi$ on system A can be realized through a unitary (Stinespring) interaction
between the system A and an environment $E$ \cite{stinespring}, i.e., 
\begin{align}
\Phi(\hat{\rho}^A) =
\rm{Tr}_E[U_{AE}\,(\hat{\rho}^A \otimes |\psi\rangle_E \langle \psi|)U_{AE}^{\dagger}],
\label{ir2}
\end{align}
$|\psi_E\rangle \in {\cal H}_E$ being a fixed pure state.
It is clear that the channel is fully specified by the joint unitary $U_{AE}$
and a fixed fiducial state $|\psi\rangle_E $. 

We can use either the complete positivity
requirement of Eq.\,(\ref{ir1}) or the unitary dilation in Eq.\,(\ref{ir2})
to obtain the operator-sum representation of a channel. In the latter situation,
the operator-sum is developed by performing the partial trace on the environment in a suitably chosen basis $\{|e_{k}\rangle_{E} \}$,
after the joint unitary evolution $U_{AE}$,
 to obtain
\begin{align}
\Phi(\hat{\rho}^A)  = \sum_{k}
[{}_{E}\langle e_{k}| U_{AE}|\psi \rangle_{E}]\,\hat{\rho}^A
\,[{}_{E}\langle \psi| U_{AE}^{\dagger}|e_{k}\rangle_{E}],
\label{ir4}
\end{align}
so that $A_{k}={}_{E}\langle e_{k}| U_{AE}|\psi \rangle_{E} $ are the Kraus operators acting on ${\cal H}_A$.
In fact, this was the procedure deployed
in\,\cite{gaussch} to derive a set of Kraus operators for {\em all} single-mode
bosonic Gaussian channels.

\section{Coherent realization of Kraus operators \label{cohrep}}
\begin{figure}
\begin{center}
\includegraphics[width=\columnwidth]{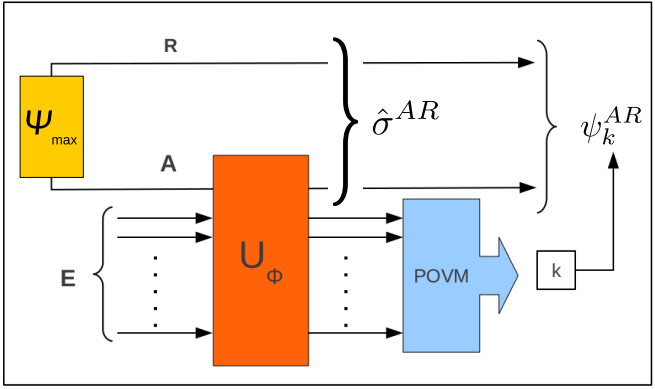}
\end{center}
\caption{Coherent realization of Kraus operators.
  The channel $\Phi$ is applied subsystem A of a bipartite maximally entangled state $\ket{\psi_{\rm max}^{AR}}$, and the channel is represented by its Stinespring dilation $U_{\Phi}$ with a suitable pure environment state E. To obtain a pure state decomposition of $\hat{\sigma}^{AR} = \frac{1}{d} \hat{\rho}^{AR}$, we apply a  suitable rank-one POVM on the environment modes E following the action by the channel unitary $U_{\Phi}$, the purification modes of $\hat{\sigma}^{AR}$. Conditioned on a measurement outcome $k$, the remaining $AR$ modes are projected into pure states $\ket{\psi_k^{AR}}$. The Kraus operators are finally obtained by  applying the Choi-Jamiolkowski isomorphism to the collection of pure states $\{\ket{\psi_k^{AR}} \}$.}
\label{fig1}
\end{figure}
The coherent representation is a way to obtain Kraus operators where all the involved mixed states are purified and the channels are dilated to its corresponding unitary as depicted in Fig. \ref{fig1}. As mentioned in Sec. \ref{pre},
it follows by Eq. \eqref{ir5a}, which we repeat for convenience, that
\begin{align}
\hr{}^{AR} = [\Phi \otimes 1\!\!1_R] (\hat{\Gamma}^{AR}) = \sum_k \ket{A_k} \bra{A_k}.
  \end{align}
  So the well known characterization of all pure state decompositions or ensemble realizations of a specific mixed  state\,\cite{hugston93}
also gives the complete enumeration of all possible operator-sum representations of a given channel.

It is useful to recall that any pure state decomposition of a given mixed state can be realized
by applying a rank-one POVM measurement on the ancillary system
of its purification\,\cite{Nielsen-Chuang2002,hugston93}. For the coherent realization of Kraus operators,
an operator-sum representation of a given channel $\Phi$ can be obtained
from the purification of $\hat{\sigma}^{AR}$ using the Stinespring dilation of the channel.
Every rank-one POVM over the output environment ancillary system $E$
(\,the purification of $\hat{\sigma}^{AR}$\,) first gives a pure state decomposition of $\hat{\sigma}^{AR}$, then using the Choi-Jamiolkowski isomorphism, we can obtain an operator-sum representation for the channel\,\cite{kraus-meas1,kraus-meas2}.

\section{Gaussian states and Gaussian channels \label{gausstuff}}
We now present some background material for Gaussian states and channels that is used in this paper. Any $N$-mode Gaussian pure state $\ket{\psi}$ can be obtained by applying a Gaussian unitary $U$
to the vacuum state $\ket{0}^{\otimes N}$, i.e., $\ket{\psi}=U\ket{0}^{\otimes N}$,
where $U$ is the exponential of a  Hamiltonian quadratic 
in the creation and annihilation operators of the $N$ modes\,\cite{simon87a,simon87b}.
Similarly, every Gaussian mixed state results from applying such a
Gaussian unitary $U$ to a tensor product
of $N$ modes in thermal states, with potentially different average number of photons (or temperatures)  \cite{dutta94a,dutta94b}.
The characteristic function $\chi({\boldsymbol \xi};\hr{})$ of a Gaussian state $\hat{\rho}$
is a Gaussian function
\begin{align}
 \chi({\boldsymbol \xi};\hr{})={\rm exp}\left(-{\boldsymbol \xi}^{\,T}V
{\boldsymbol \xi}/2-i{\boldsymbol d}^T\Omega {\boldsymbol \xi}\,\right),
\end{align}
specified completely  by the mean ${\boldsymbol d}$ and variance matrix $V$.
Here ${\boldsymbol \xi}$ represents a point in the N-mode phase space viewed as a column
vector ${\boldsymbol \xi}=(q_1,p_1,q_2,p_2,\cdots,q_n,p_n)^T$.
The positivity of the Gaussian state $\hat{\rho}$ is equivalent to
the uncertainty relation\,\cite{simon87a,simon87b,dutta94a,dutta94b}
\begin{align}
 V+i\Omega\geq0, \,\, \,\, \Omega=\bigoplus_{i=1}^N
\left(\begin{matrix}
0 & 1 \\
-1 & 0
\end{matrix}\right),
\label{gur}
\end{align}
which is the multimode generalization of the Schr\"{o}dinger-Robertson uncertainty principle in the single mode case.
A Gaussian unitary transformation $U$ results in a homogeneous linear phase space transformation ${\boldsymbol \xi} \to S^T{\boldsymbol \xi}$
and a phase space displacement ${\boldsymbol \xi} \to {\boldsymbol \xi} + {\boldsymbol d}$.
Here the symplectic matrix $S$ (satisfying $S\Omega S^T=\Omega$)
takes $ \chi({\boldsymbol \xi};\hr{}) \to\chi(S^T{\boldsymbol \xi};\hr{})$,
so that the variance matrix $V$ transforms as $V \to SVS^T$ and the displacement ${\bs d} \to S^T{\bs d}$\,\cite{simon87a,simon87b,dutta94a,dutta94b}.

Gaussian channels are quantum processes that map input Gaussian states to Gaussian states at the output.  Gaussian channels are characterized by a pair of real matrices,
a  transmission matrix $X$ and a symmetric noise matrix $Y\geq 0$.
To guarantee that a given pair $(X, Y)$ represents a completely positive trace preserving map, it has to satisfy the necessary and sufficient condition\,\cite{cp1,cp2,cp3}
\begin{align}
 Y+i\Omega-iX^T\Omega X\geq0.
\end{align}
The action of the Gaussian channel $\Phi(X,Y)$ is conveniently described at the level of the characteristic function as
\begin{align}
 \chi({\boldsymbol \xi};\Phi[\hr{}])  = \chi(X\, {\boldsymbol \xi};\hr{})\,
\exp\left({-\frac{1}{2}{\boldsymbol \xi}^{\,T}\, Y \, {\boldsymbol \xi}}\right).\nonumber\\
\end{align}
It follows that the variance matrix of a state under the action of a Gaussian channel $\Phi(X,Y)$ is governed by the simple  relation
\begin{align}
 V'=X^T\,V\,X +Y.
\label{gact}
\end{align}
Further, a Gaussian channel is said to be entanglement breaking \cite{holevo08} if its corresponding noise matrix $Y$ can be decomposed into noise matrices $Y = Y_1 + Y_2$ such that
\begin{align}
Y_1 \geq i\Omega, ~~ Y_2 \geq i\, X^T \Omega X.
  \end{align}
In the simplest case of single-mode Gaussian channels
the matrices $(X,Y)$ are  $2\times 2$, and the complete-positivity condition
can be simplified to the scalar condition
\begin{align}
 \det Y\geq \left(\det X-1\right)^2.
\label{eq:CPTPcondition}
\end{align}

The canonical channels were completely characterized in Refs.\,\cite{holevo07a,holevo07b}. The channels  with non-singular $X$,  are more important and can be divided into two different classes based on the signature of ${\rm det}\,X$.  First, the attenuator/amplification channels
with $X_0= \kappa 1\!\!1$ ($\kappa > 0$), where $\kappa^2$ is the loss or gain of the channel
and noise $Y_0=\alpha 1\!\!1$ with  $\alpha \geq|\kappa^2-1|$ (resulting from Eq.\,(\ref{eq:CPTPcondition}));
the attenuator channel is denoted by ${\cal C}_{1}(\kappa,\, \alpha)$ with $\kappa \leq 1$,
and the amplification channel by ${\cal C}_{2}(\kappa,\, \alpha)$ with $\kappa \geq 1$.
The second class comprises the phase conjugation channel denoted  ${\cal D}(\kappa,\,\alpha)$,
with $X_0=\kappa \sigma_3$ and  noise $Y_0=\alpha 1\!\!1$. Here $\kappa > 0$
with $\alpha \geq \kappa^2+1$, and $\sigma_3$ is the Pauli matrix.
The classical noise channel ${\cal B}_{2}(\alpha)$ is obtained as
the limit $\kappa \to 1$ as in ${\cal C}_{1}(1,\,\alpha)={\cal C}_{2}(1,\,\alpha)$ of {\em either} the noisy attenuator {\em or}
noisy amplifier channel.
To complete the presentation of the one-mode canonical channels we mention
the singular channel ${\cal A}_{2}(\alpha)$ for which
$X_0=(1\!\!1+\sigma_3)/2$ and $Y_0=\alpha 1\!\!1$ with $\alpha\geq1$, the
single quadrature classical noise channel ${\cal B}_{1}(\alpha)$
for which $X_0=1\!\!1$, and $Y_0= \alpha(1\!\!1+\sigma_3)/2$ with $\alpha \geq 0$, and the full loss channel ${\cal A}_1(\alpha)$ with $(X_0,Y_0) = (0,\alpha 1\!\!1)$, and $\alpha \geq 1$.

A single-mode Gaussian channel is said to be quantum-limited if the Gaussian noise $\alpha$
is no larger than the minimum required to saturate the completely positivity condition in  Eq. (\ref{eq:CPTPcondition}). 
That any non quantum-limited Gaussian channel
can be obtained as concatenation of a pair of quantum-limited
Gaussian channels is proved in\,\cite{gaussch,garciapatron12}.
Finally, all quantum-limited Gaussian channels are extremal and these are the only extremal Gaussian channels \cite{gaussch,holevoext}.

\section{Quantum-limited Amplifier ${\cal C}_2(\kappa)$ \label{qlamp}}
We treat the quantum-limited amplifier $\mathcal{C}_2(\kappa)$ in an analogous manner to the quantum-limited attenuator channel,  so that the
requirement of Eq.\,\eqref{b2b} reduces to ${\rm tanh}\,\mu = \kappa^{-1} \,{\rm tanh}\,r $.
The symplectic transformation in Eq.\,\eqref{b2a} of step 2  is given by
\begin{align}
S[\mu(r, \kappa)] = \frac{1}{\sqrt{\kappa^2 -\tanh^2{r}}}\left[
\begin{array}{cc}
\kappa\,1\!\!1_2 & - \tanh{r}\, \sigma_3\\
- \tanh{r}\, \sigma_3 & \kappa\,1\!\!1_2
\end{array}
\right].
\label{a0}
\end{align}
In the third step we find the symplectic eigenvalues 
  $\nu^{+}_{a}=(\kappa^2-1) \cosh{2r}+\kappa^2$ and $\nu^{-}_{a}=1$.
So $\hat{\sigma}_{\rm can}^{a}(r, \kappa)$ is a product of a
thermal state in mode 1 and the ground state in mode 2. Written in the Fock basis we have
\begin{align}\label{a1}
\hat{\sigma}_{\rm can}^{a}(r,\kappa) &={\rm sech}^2 r\,N_{a}(r, \kappa)\sum_{n=0}^{\infty}
  (x(\nu^{+}_{a}))^{2n} |n,0 \rangle \langle n, 0 |,
\end{align}
with
\begin{align}
N_{a}(r,\kappa) &= [\kappa^2 - \tanh^2r]^{-1},\nonumber\\
x(\nu_a^+) &= \left[\frac{\nu_a^+-1}{\nu_a^++1}\right]^{1/2}= \left[\frac{\kappa^2-1}{\kappa^2-\tanh^2{r}}\right]^{1/2}.
\label{a1b}
\end{align}
In step 4, we  undo the unitary two-mode squeezing transformation of Eq.\,\eqref{a0}
on $\hat{\sigma}^a_{\rm can}(r,\kappa)$ to obtain
\begin{align}\label{a2}
\hat{\sigma}^{a}(r,\kappa) &=  U_S^{\dagger}\,\hat{\sigma}_{\rm can}^{a}(r,\kappa) \,U_S = {\rm sech}^2 r \, \hat{\rho}^a(r,\kappa),  \nonumber\\
\hat{\rho}^a(r,\kappa)  &=\sum_{n=0}^{\infty}\proj{{T}_{n}^{a}(r,\kappa)},
\end{align}
where
\begin{align}
|T_{n}^{a}(r,\kappa)\rangle &= (x(\nu^{+}_{a}))^n \sqrt{N_{a}(r,\kappa)}\,U_S^{\dagger}|n, 0 \rangle \nonumber\\
&= (x(\nu^{+}_{a}))^n \sqrt{N_{a}(r,\kappa)}\nonumber\\
& \times  \sum_{m=0}^{\infty} \sqrt{\binom{m+n}{n}}\,
\frac{\tanh^{m}{\mu}}{\cosh^{n+1}\mu}|m+n,m \rangle.
\label{a3}
\end{align}
The Kraus operators are
\begin{align}\label{a4}
{T}_{n}^{a}(r,\kappa) &= \sqrt{N_{a}(r,\kappa)} \, (x(\nu^{+}_{a}))^n  \nonumber \\
& \times \sum_{m=0}^{\infty} \sqrt{\binom{m+n}{n}}\,
\frac{\tanh^{m}{\mu}}{{\rm cosh}^{n+1}{\mu}} |m+n\rangle \langle m|.
\end{align}
As  with ${\cal C}_{1}(\kappa)$, the Kraus operators $\{{T}_{n}^{a}(r, \kappa) \}$ represent
a completely positive map ${\cal C}_{2}(r, \kappa)$ which is not trace-preserving.  
Substituting the expressions for $N_a(r,\kappa),\,x(\nu_a^{+})$ from Eq.\,\eqref{a1b} and for  $\tanh{\mu},\,\cosh{\mu}$ from Eq.\,\eqref{a0}, we obtain
\begin{align}
\left(\sum_{n=0}^{\infty} T^{a}_n(r,\kappa)^{\dagger} T^a_n(r,\kappa)\right)_{j \ell} = \delta_{j\ell} \,(\tanh^2{r})^j.
\label{atp}
\end{align}
Returning to Eq.\,\eqref{a4}, in the limit $r \to \infty$ we have $x(\nu^{+}_{a}) \to 1$,  $N_{a}(r,\kappa) \to ({\kappa^2 -1})^{-1}$,
${\rm  tanh}[2\,\mu(r,\kappa)] \rightarrow {\rm  tanh}[2\,\mu_0(\kappa)] ={2\kappa}/({1+\kappa^2})$,
${\rm sech}[\mu(r,\kappa)]\rightarrow {\rm sech}[\mu_0(\kappa)]
={\sqrt{ \kappa^2-1}}/{\kappa}$, and
${\rm tanh}[\mu(r,\kappa)]\rightarrow {\rm tanh}[\mu_0(\kappa)]={\kappa}^{-1}$.
Substituting these limits, we have the final expression for the Kraus operators of the quantum-limited
amplifier channel to be
\begin{align}\label{a5}
&A_n(\kappa)=\lim_{r \to \infty} T^a_n(r,\,\kappa)\nonumber\\
&=\sum_{m=0}^{\infty} \sqrt{\binom{m+n}{n}}
\left(\frac{1}{\kappa}\right)^{m+1}\!\! \left(\frac{\sqrt{\kappa^2-1 }}{\kappa}\right)^{n}
\!\!|m+n \rangle \langle m|, \nonumber \\
&\hspace{3cm}\,\,n=0,1,2,\cdots.
\end{align}
These Kraus operators are seen to restore the expected trace-preserving property
\begin{align}
\lim_{r \to \infty} \,\sum_{n=0}^{\infty} T_n^a(r,\kappa)^{\dagger} T^a_n(r,\kappa) = \sum_{n=0}^{\infty} A_n(\kappa)^{\dagger} A_n(\kappa) =1\!\!1.
\end{align}

\section{Phase conjugation channel \label{qlpc}}
In addition to the attenuator and amplifier channels presented earlier there exists a third class of channels characterized by nonsingular $X$, with ${\det X}<0$;
these are the phase conjugation channels  $D(\kappa,\alpha)$ specified (in its canonical form)
by $(X,Y)=(\kappa \sigma_3, \alpha 1\!\!1)$,  $\alpha\geq\kappa^2+1$.
Clearly, the variance matrix of the output bipartite state
resulting from the one-sided action of
the channel on the canonical two-mode squeezed vacuum state $|\Psi_r\rangle$ in step 1 of our procedure is
\begin{align}\label{pc1}
V^c(r,\kappa,\alpha) = \left(
\begin{matrix}
(\kappa^2 \cosh{2r} +\alpha) 1\!\!1_2 &  \kappa \sinh{2r}\,1\!\!1_2  \\
\kappa \sinh{2r}\,1\!\!1_2 & \cosh{2r}\,1\!\!1_2
\end{matrix}
\right).
\end{align}
Here the superscript `$c$' is meant to remind us  that we are
working with the phase conjugation channel.
It is clear that $(\kappa\sigma_3, \alpha1\!\!1)$ represents a channel  only
if both the symplectic eigenvalues $\nu^{+}_{c}$, $\nu^{-}_{c}$ $\geq1$ for
all $r$, but this is readily seen to hold  if and only if $\alpha \geq \kappa^2+1$, {\em independent of $r$}.
Since the determinant of the off-diagonal block of $V^c(r,\kappa,\alpha)$
is always positive, we know by the little lemma of Ref.\,\cite{simon00} that
the output Gaussian state is always separable, {\em irrespective of the value of $r$}.
Equivalently, the requirement that the partial transpose
$V^c(r,\kappa,\alpha)^{PT}$ satisfies the uncertainty principle, is weaker
than that of the uncertainty principle on $V^{c}(r,\kappa,\alpha)$.
Therefore, every phase conjugation channel is automatically entanglement breaking irrespective of the values of the channel parameters.

Additionally, since $V^c(r,\kappa,\alpha)$ 
has all {\em $2 \times 2$ blocks proportional to identity}, it can be diagonalized by the beam splitter
symplectic transformation (equal rotations in the $q_{1}-q_{2}$ and $p_1-p_{2}$ planes)
\begin{align}
S(\theta(r,\kappa,\alpha)) = \left( \begin{array}{cc}
\cos{\theta}\, 1\!\!1_2 & -\sin{\theta}\, 1\!\!1_2\\
\sin{\theta}\, 1\!\!1_2 & \cos{\theta}\, 1\!\!1_2
\end{array} \right),
\label{pc2}
\end{align}
where
\begin{align}
\tan\left[{2\,\theta}(r,\kappa, \alpha)\right]= \frac{2\kappa \sinh{2r}}{\cosh{2r}-\kappa^2\cosh{2r}-\alpha}.
\label{pc3}
\end{align}
Thus, the (doubly degenerate) symplectic eigenvalues of $V^c(r,\kappa,\alpha)$ are
\begin{align}\label{pc4}
\nu^{\pm}_{c}&=\frac{1}{2} \left[w_{c} \pm z_{c}\right],\nonumber \\
z_c&=[(\alpha+(\kappa^2-1)\cosh{2r})^2+4\kappa^2\sinh^2{2r}]^{1/2}, \nonumber \\
w_c&=\alpha+\cosh{2r}\kappa^2+\cosh{2r}.
\end{align}

\subsection{Quantum-limited phase conjugation channel ${\cal D}(\kappa)$}
For the quantum-limited phase conjugation channel ($\alpha =\kappa^2 +1$)
we obtain at the end of step 2, i.e., after the symplectic diagonalization,
one mode in vacuum ($\nu^{-}_{c}=1$) with the other mode in a thermal state
corresponding to the symplectic eigenvalue $\nu^{+}_{c}=\cosh{2r}(\kappa^2+1)+\kappa^2$,
 and Eq.\,\eqref{pc3} reduces to ${\rm tan}\,\theta = -{\rm tanh}\,r/\kappa$.
The symplectic transformation of Eq.\,\eqref{pc2} of step 2 that leads to
the diagonalization of $V^c(\kappa,r)$ is
\begin{align}
S[\theta(r, \kappa)] = \frac{1}{\sqrt{\kappa^2 +\tanh^2{r}}}\left(
\begin{array}{cc}
\kappa\,1\!\!1_2 &  \tanh{r}\,1\!\!1_2\\
- \tanh{r}\,\,1\!\!1_2 & \kappa\,1\!\!1_2
\end{array}
\right).
\label{pc4b}
\end{align}
In step 3, $\hat{\sigma}_{\rm can}^{c}(r, \kappa)$ is decomposed in the Fock basis as
\begin{align}
\hat{\sigma}_{\rm can}^{c}(r, \kappa)&={\rm sech}^2 r \,N_{c}(r,\kappa) \sum_{n=0}^{\infty}
(x(\nu^{+}_{c}))^{2n} | n, 0 \rangle \langle n, 0 |,
\label{pc5}
\end{align}
where the scalar factor
\begin{align}
N_{c}(r, \kappa)=[1+\kappa^2]^{-1}
\label{pc50}
\end{align} is {\em independent} of $r$, and
\begin{align}
x(\nu^{+}_{c}) = \left[\frac{\nu^{+}_{c}-1}{\nu^{+}_{c}+1}\right]^{1/2}=\left[\frac{\kappa^2 + \tanh^2{r}}{\kappa^2+1}\right]^{1/2}.
\label{pc51}
\end{align}
In step 4, we  undo the (beam-splitter) unitary transformation
on $\hat{\sigma}_{\rm can}^{c}(r,\kappa)$ to obtain
\begin{align}
  \hat{\sigma}^{c}(r,\kappa) &= {\rm sech}^2 r \,\hat{\rho}^c(r,\kappa),\nonumber\\
\hat{\rho}^c(r,\kappa) &=  \sum_{n=0}^{\infty}\proj{{T}_{n}^{c}(r,\kappa)},
\label{pc5b}
\end{align}
where
\begin{align}
|{T}_{n}^{c}(r,\kappa)\rangle &=
\sqrt{N_{c}(r, \kappa)} (x(\nu^{+}_{c}))^n \,U_S^{\dagger}\, |n, 0 \rangle \nonumber \\
&= \sqrt{N_{c}(r, \kappa)} (x(\nu^{+}_{c}))^n \nonumber\\
&\times \sum_{m=0}^{n}\sqrt{\binom{n}{m}}\left(-\sin{\theta}\right)^{m}
\left(\cos{\theta}\right)^{n-m}\ket{n-m,m}.
\label{pc6}
\end{align}
We  read off the Kraus operators ${T}_{n}^{c} (r,\kappa)$ from the bipartite vectors $|T_n^c(r,\kappa)\rangle$ as
\begin{align}\label{pc7}
&{T}_{n}^{c}(r,\kappa) = \sqrt{N_{c}(r,\kappa)} \, (x(\nu^{+}_{c}))^n  \nonumber \\
& \times \sum_{m=0}^{n}\sqrt{\binom{n}{m}}\left(- \sin{\theta}\right)^{m}
\left(\cos{\theta}\right)^{n-m}\ket{n-m} \langle m|.
\end{align}
 The map  ${\cal D}(r,\kappa)$ resulting from the Kraus operators
$\{{T}_{n}^{c}(r, \kappa) \}$ is not trace-preserving for finite $r$. Indeed 
 by Eqs.\,\eqref{pc4b}, \eqref{pc50} and \eqref{pc51}, we find that
\begin{align}
\left(\sum_{n=0}^{\infty} T^{c}_n(r,\kappa)^{\dagger} T^c_n(r,\kappa)\right)_{j \ell} = \delta_{j\ell} \,(\tanh^2{r})^j.
\end{align}
Going back to Eq.\,\eqref{pc7},
in the limit $r \rightarrow \infty$
 we have $x(\nu^{+}_{c}) \to 1$,
$\sin[\theta(r,\kappa)] \rightarrow \sin[\theta_0(\kappa)]={-1}/{\sqrt{1+\kappa^2}}$,
$\cos[\theta(r,\kappa)] \rightarrow \cos[\theta_0(\kappa)] ={\kappa}/{\sqrt{1+\kappa^2}}$,
and $\tan[\theta(r,\kappa)] \rightarrow \tan{[\theta_0(\kappa)]}=-{\kappa}^{-1}$, where $\lim\limits_{r \to \infty} \theta(r,\kappa) = \theta_0(\kappa)$.
Substituting these limits in Eq.\,\eqref{pc7}, the Kraus operators of the
quantum-limited phase conjugation channel are
\begin{align}
C_{n}(\kappa) &= \lim_{r \to \infty} T^c_n(r,\,\kappa)\nonumber\\
&=\frac{1}{\sqrt{1+\kappa^2}}
\sum_{m=0}^{n}\sqrt{\binom{n}{m}} \,\left[\frac{\kappa}{\sqrt{1+\kappa^2}}\right]^{n-m}
 \nonumber \\
& \times \left[\frac{1}{\sqrt{1+\kappa^2}} \right]^m  \,
|n-m \rangle \langle m|,\, n=0,1,\cdots,~~~~~~~~
\label{pc8}
\end{align}
and  restore the `resolution of identity' 
\begin{align}
\lim_{r \to \infty} \,\sum_{n=0}^{\infty} T_n^c(r,\kappa)^{\dagger} T^c_n(r,\kappa) = \sum_{n=0}^{\infty} C_n(\kappa)^{\dagger} C_n(\kappa) =1\!\!1.
\end{align}

\section{Noisy Gaussian channels \label{noisyc}}
In this section we derive an operator-sum representation for noisy attenuator, amplifier, and phase conjugation channels. A channel is said to be noisy when its associated noise matrix $Y$ is `larger' than the corresponding value for the quantum-limited case. We compare the Kraus operators with another set obtained  from the concatenation of the quantum-limited channels detailed in Ref. \cite{gaussch}. 

\subsection{Noisy attenuator and amplifier channels \label{naa}}
In the noisy cases of both the attenuator and amplifier channels, we no longer have in step 2 of our procedure  outlined in Sec. \ref{tech} the luxury of one of the symplectic eigenvalues evaluating to unity corresponding to the vacuum state. So, $\hat{\sigma}^{b/a}_{\rm can}(r)$ in step 2
now becomes a product of two thermal states. In step 3, $\hat{\sigma}^{b/a}_{\rm can}(r)$ is expanded in the Fock basis as a {\em double sum} rather than a single-index sum. Undoing the
unitary $U_{S}$ in step 4, we obtain $\hat{\sigma}(r)$ to be a
double indexed convex sum of rank-one bipartite projections.  As a consequence, each Kraus operator is
now labeled by a pair of indices as opposed to a single index in the quantum-limited situation.

The symplectic eigenvalues of $V^{b/a}_{\rm can}(r,\kappa,\alpha)$ for noisy ${\cal C}_1(\kappa,\alpha)$ and ${\cal C}_2(\kappa,\alpha)$
are given in Eq.\,(\ref{b4}) that now depend on the noise parameter $\alpha$.
During step 3, the product of thermal states corresponding to $V^{b/a}_{\rm can}(r,\kappa,\alpha)$ is decomposed
in the Fock basis as
\begin{align}
\hat{\sigma}^j_{\rm can}(r,\kappa,\alpha) &= {\rm sech}^2{r}\,N_j(r,\kappa,\alpha)  \sum_{n_1,n_2=0}^{\infty} (x(\nu_j^+))^{2n_1} \nonumber\\ & ~~~~ \times (x(\nu_j^-))^{2n_2}
|n_1,n_2\rangle \langle n_1,n_2|,\nonumber\\
N_j(r,\kappa,\alpha)  &= \cosh^2{r}\,\left(1-(x(\nu_j^+))^2 \right) \left(1-(x(\nu_j^-))^2\right),
\label{ap1}
\end{align}
where the index $j$ stands for either $a$ or $b$ according to an amplifier or beam splitter channel.
In step 4, we undo the two-mode squeeze transformation to obtain
\begin{align}
&\hat{\sigma}(r,\kappa,\alpha)= U_S[\mu]^{\dagger}\,\hat{\sigma}_{\rm can}(r,\kappa,\alpha) \,U_S[\mu] \nonumber\\
&=  {\rm sech}^2{r} \sum_{n_1,n_2=0}^{\infty}
|T^j_{n_1,n_2}(r,\kappa,\alpha) \rangle \langle T^j_{n_1,n_2}(r,\kappa,\alpha)|,
\end{align}
where
\begin{align}
|T^j_{n_1,n_2}(r,\kappa,\alpha) \rangle = \sqrt{N_j(r,\kappa,\alpha)} \, (x(\nu_j^+))^{n_1} (x(\nu_j^-))^{n_2} \nonumber\\ ~~~~~~~\times  \,U_S^{\dag}[\mu] |n_1,n_2\rangle.
\end{align}
Using the Fock basis matrix elements $\langle m_1, m_2|U_S[\mu]^{\dagger}|n_1,n_2\rangle$ of the two-mode squeeze operator $U_S[\mu]^{\dagger}$ (see Eq. (5.4) of \cite{gaussch}) we have the pure states to be
\begin{align}
&|T^j_{n_1,n_2}(r,\kappa,\alpha) \rangle = \sum_{m_1={\rm max}\,\{0,n_1-n_2\}}^{\infty} \hspace{-0.6cm} \gamma_j[\mu] \, |m_1,n_2+m_1-n_1\rangle, \nonumber\\
& \gamma_j[\mu] = \sqrt{N_j(r,\kappa,\alpha)} \, (x(\nu_j^+))^{n_1}\, (x(\nu_j^-))^{n_2}\, h_j[\mu],\nonumber\\
  &h_j[\mu] = \sum_{r=\,{\rm max}\,\{0,n_1-m_1\}}^{{\rm min}\,\{n_1,n_2\}} \left[\frac{n_1! (n_2+m_1-n_1)!}{n_2! m_1!}\right]^{1/2} \binom{n_2}{r} \nonumber\\ &\times \,\binom{m_1}{n_1-r}\,(-1)^r\,({\rm sech}{\mu})^{n_1+n_2-2r+1} \,(\tanh{\mu})^{2r + m_1 -n_1}.
  \label{absm0}
\end{align}
We now associate the Kraus operators $T^j_{n_1,n_2}$ with  unnormalized bipartite states $|T^j_{n_1,n_2}\rangle$ by flipping the second ket to a bra\,:
\begin{align}
&T^j_{n_1,n_2}(r,\kappa,\alpha)  = \sum_{m_1={\rm max}\,\{0,n_1-n_2\}}^{\infty} \hspace{-0.7cm} \gamma_j[\mu] \, \ket{m_1}\bra{n_2+m_1-n_1}.
\label{asbm1}
\end{align}
Now consider the noisy attenuator channel\,($\kappa<1$ and $\alpha > 1-\kappa^2$) and the index $j=b$.
In the limit $r \rightarrow \infty$ we see that $\nu^{-}_{b} \rightarrow
\infty \Rightarrow  x(\nu_b^-) \to  1$, while $\nu^{+}_{b} \rightarrow \alpha(1-\kappa^2)^{-1}$, and
$\cosh{\mu}$, $\sinh{\mu}$ have the same limits as in the quantum-limited case. So we obtain from Eq. \eqref{absm0} 
\begin{align}
\gamma_b[\mu_0] &= \left[ \frac{2}{\alpha+1-\kappa^2}\right]^{1/2}\, \left[ \,\frac{\alpha + \kappa^2 -1}{\alpha+1-\kappa^2}\right]^{n_1/2}\, h_b[\mu_0].\nonumber\\
\label{asbm2}
\end{align}
The final Kraus operators of the noisy attenuator channel are read off from Eqs.\,\eqref{asbm1} and \eqref{asbm2} as
\begin{align}\label{pp1}
  B_{n_1,n_2}(\kappa,\alpha) &=\sum_{m_1={\rm max}\,\{0,n_1-n_2\}}^{\infty} \hspace{-0.4cm} \gamma_b[\mu_0] \, \ket{m_1}\bra{n_2+m_1-n_1}.
%
\end{align}

For the noisy amplifier channel $\kappa>1$ and $\alpha > \kappa^2-1$ and $j=a$.
In the limit $r \rightarrow \infty$ we see that $\nu^{+}_{a} \rightarrow
\infty \Rightarrow  x(\nu_a^+) \to  1$, while $\nu^{-}_{a} \rightarrow \alpha(\kappa^2-1)^{-1}$, and
$\cosh{\mu}$, $\sinh{\mu}$ have the same limits as in the quantum-limited case. So  we then obtain 
\begin{align}
\gamma_a[\mu_0] &= \left(\frac{2}{\alpha +  \kappa^2 -1 }\right)^{1/2} \, \left( \frac{\alpha - \kappa^2 +1}{\alpha+\kappa^2-1}\right)^{n_2/2}\,   h_a[\mu_0].\nonumber\\
\label{asbm3}
\end{align}
Again, the final Kraus operators of the noisy amplifier channel are read off from Eqs.\,\eqref{asbm1} and \eqref{asbm3} as
\begin{align}\label{pp2}
  A_{n_1,n_2}(\kappa,\alpha) &= \sum_{m_1={\rm max}\,\{0,n_1-n_2\}}^{\infty} \hspace{-0.4cm} \gamma_a[\mu_0] \, \ket{m_1}\bra{n_2+m_1-n_1}.
\end{align}

\noindent
{\bf Remark.} It is well-known that the operator-sum representation of a channel is not unique, and two different sets
of Kraus operators can represent one and the same channel. In the case of
a noisy ${\cal C}_1(\kappa,\alpha)$ or ${\cal C}_2(\kappa,\alpha)$,
one can  obtain an alternative set of Kraus
operators using the fact that the noisy channel can be realized\,\cite{gaussch,holevo07a,garciapatron12} as product
${\cal C}_2(\kappa_2) \circ {\cal C}_1(\kappa_1)$
of a quantum-limited attenuator with transmissivity $T = \kappa_1^2$ followed by a quantum-limited amplifier with gain $G = \kappa_2^2$.
The resulting channel is a noisy ${\cal C}_1(\kappa,\alpha)$ or ${\cal
  C}_2(\kappa,\alpha)$, depending on whether the product
$\kappa_1\kappa_2$ is $< 1$ or $> 1$, with $\kappa=\kappa_1 \kappa_2$ and
$\alpha=\kappa_2^2(1-\kappa_1^2)+\kappa_2^2-1$ in both cases.

Now consider the product ${\cal C}_{2}(\kappa_2)\circ {\cal C}_{1}(\kappa_1)$. By choosing the special value $\kappa_1 \kappa_2 =1$ one obtains ${\cal B}_{2}(\alpha)$
with $\alpha =2(\kappa_{2}^{2}-1)$, which is the classical noise
channel. The double-index {\em discrete sum} Kraus representation
resulting from ${\cal C}_2(\kappa_2) \circ {\cal C}_1(\kappa_1)$  is an {\em alternative} to the more
familiar continuous sum Kraus representation in terms of phase space
displacement operators $ D(\beta) = \exp[\beta a^{\dagger} - \beta^* a]$\,\cite{hall94}.
It is clear that one
obtains, in general,  a set of double-indexed Kraus operators for the noisy channel ${\cal C}_{1}(\kappa,\alpha)$
or ${\cal C}_{2}(\kappa,\alpha)$ as simply products
of the two sets of the single-indexed Kraus operators $\{B_i(\kappa_1) \}_i$ and $\{A_j(\kappa_2) \}_j$, i.e.,  $T_{i,j}(\kappa,\alpha)= B_i(\kappa_2)\,A_j(\kappa_1)$, where $i,j$ are indices to label the Kraus operators  obtained using an  entirely different point of view \cite{gaussch}.\hfill $\blacksquare$

\subsection{Noisy phase conjugation channel ${\cal D}(\kappa, \alpha)$}
The method for obtaining Kraus operators for {\em noisy} phase conjugation
channels  is similar to that of the noisy attenuator and amplifier
channels. 
 As in the earlier two cases, during step 3 the product thermal state
$\hat{\sigma}_{\rm can}^c(r,\kappa,\alpha)$
corresponding to $V_{\rm can}(r,\kappa,\alpha)$ is decomposed in the Fock basis as
in Eq.\,(\ref{ap1}) with the symplectic eigenvalues given as in Eq.\,(\ref{pc4}).

In step 4 we undo the beam splitter unitary to obtain
\begin{align}
&\hat{\sigma}^c(r,\kappa,\alpha)= U_S[\theta]^{\dagger}\,\hat{\sigma}_{\rm can}(r,\kappa,\alpha) \,U_S[\theta] \nonumber\\
&=  {\rm sech}^2{r} \sum_{n_1,n_2=0}^{\infty}
|T^c_{n_1,n_2}(r,\kappa,\alpha) \rangle \langle T^c_{n_1,n_2}(r,\kappa,\alpha)|,
\end{align}
where
\begin{align}
&|T^c_{n_1,n_2}(r,\kappa,\alpha) \rangle \nonumber \\&= \sqrt{N_c(r,\kappa,\alpha)} \, (x(\nu_c^+))^{n_1} \,(x(\nu_c^-))^{n_2} \,U_S[\theta]^{\dag} |n_1,n_2\rangle,\nonumber\\
&N_c(r,\kappa,\alpha)  = \cosh^2{r}\,\left(1-(x(\nu_c^+))^2 \right) \,\left(1-(x(\nu_c^-))^2\right).
\end{align}
Using the Fock basis matrix elements $\langle m_1,m_2|U_S[\theta]^{\dagger}|n_1,n_2\rangle$ of the beam splitter unitary $U_S^{\dagger}[\theta]$ (see Eq. (4.4) of \cite{gaussch}) we obtain
\begin{align}
&|T^c_{n_1,n_2}(r,\kappa,\alpha) \rangle = \sum_{m_2=0}^{n_1+n_2} \gamma_c[\theta] \, |n_1+n_2-m_2,m_2\rangle, \nonumber\\ 
 &\gamma_c[\theta]  = \sqrt{N_c(r,\kappa,\alpha)} \, (x(\nu_c+))^{n_1}\, (x(\nu_c^-))^{n_2}\, h_c[\theta],
\end{align}
\begin{widetext}
\begin{align}
h_c[\theta] = \sum_{r=\,{\rm max}\,(0,m_2-n_2)}^{{\rm min}\,(n_1,m_2)} \left[\frac{(n_1+n_2-m_2)! m_2!}{n_1! n_2!}\right]^{1/2}  \binom{n_1}{r}\,\binom{n_2}{m_2-r}\,(-1)^{n_2-m_2+r} \,(\sin{\theta})^{2r + n_2-m_2} \,(\cos{\theta})^{n_1+m_2-2r}.
\end{align}
\end{widetext}
We now associate the Kraus operators $T^c_{n_1,n_2}$ with the unnormalized bipartite states $|T^j_{n_1,n_2}\rangle$ by flipping the second ket to a bra, i.e.,
\begin{align}
T^c_{n_1,n_2}(r,\kappa,\alpha) &= \sum_{m_2=0}^{n_1+n_2} \gamma_c[\theta] \, |n_1+n_2-m_2 \rangle \langle m_2|.
\end{align}
For the noisy phase conjugation channels  $\alpha > \kappa^2+1$.
In limit $r \rightarrow \infty$ we see that $\nu^{+}_{c} \rightarrow
\infty \Rightarrow  x(\nu_c^+) \to  1$, while $\nu^{-}_{c} \rightarrow \alpha(\kappa^2+1)^{-1}$, and
$\cos{\theta}$, $\sin{\theta}$ have the same limits as in the quantum-limited case. So putting these facts together we have
\begin{align}
\gamma_c[\theta_0] &= \left(\frac{2}{\alpha +\kappa^2+1}\right)^{1/2} \left(\frac{\alpha-\kappa^2-1}{\alpha+\kappa^2+1}\right)^{n_2/2} \,h_c[\theta_0].
\end{align}
So the final Kraus operators of the noisy amplifier channel are read off from Eqs.\,\eqref{asbm1} and \eqref{asbm3} as
\begin{align}\label{pp3}
C_{n_1,n_2}(\kappa,\alpha) &= \sum_{m_1=0}^{n} \gamma_c[\theta_0] \, |n_1+n_2-m_2 \rangle \langle m_2|.
\end{align}
\noindent
{\bf Remark.} As an alternative to the result presented here,
one can build a different set of Kraus operators for noisy phase conjugation channels using
the decomposition of a noisy phase conjugation channel of parameters
($\kappa,\alpha$) into a quantum-limited attenuator (or amplifier) channel
of transmissivity $T=\kappa_1^2$ (or gain $G= \kappa_2^2 $) followed by a quantum-limited phase conjugation
channel of parameter $\kappa^{\,'}$ and Kraus operators $\{C_n(\kappa^{\,'}) \}$.
If these channel parameters $\kappa_1$ (or $\kappa_2$) and $\kappa^{\,'}$ are arranged to
satisfy $\kappa=\kappa_1\kappa^{\,'}$ (or $\kappa=\kappa_2\kappa^{\,'}$) and $\alpha=(\kappa^{\,'})^2(2-\kappa_1^2)+1$ (or $\alpha=(\kappa^{\,'})^2\kappa_2^2+1$),
we indeed realize ${\cal D}(\kappa, \alpha)$
with Kraus operators given by the products
$T_{n,j}(\kappa,\alpha)=C_n(\kappa^{\,'})\,B_j(\kappa_1)$ [or $C_n(\kappa^{\,'})\,A_j(\kappa_1)$]. In this way one recovers the
Kraus operators presented in\,\cite{gaussch}.\,\hfill $\blacksquare$

\subsection{Connecting operator-sum representations}
We have presented a technique for developing Kraus operators
of single-mode Gaussian channels, both in the Fock and coherent state basis, exploiting the Choi-Jamiolkowski
isomorphism between completely positive maps and bipartite quantum states.
We used the fact that every decomposition of the Jamiolkowski state into pure states gives a
family of Kraus operators representing the channel. As mentioned in Appendix \ref{cohrep},
it is a well known fact that any ensemble decomposition of a given mixed state $\hat{\rho}$ is realized
by applying the appropriate rank-one POVM measurement on the ancillary system
of its purification\,\cite{Nielsen-Chuang2002,hugston93}.
Therefore, any Kraus decomposition of a given channel can be obtained by purifying
the corresponding Choi-Jamiolkowski state and applying the right rank-one POVM
to the purification or environment modes E, as show in Fig.\,\ref{fig1}.
We will now use this property to connect the operator-sum representations
obtained in\,\cite{gaussch} with the ones developed in this manuscript for noisy bosonic Gaussian channels.

\begin{figure}
\fbox{\includegraphics[width=\columnwidth]{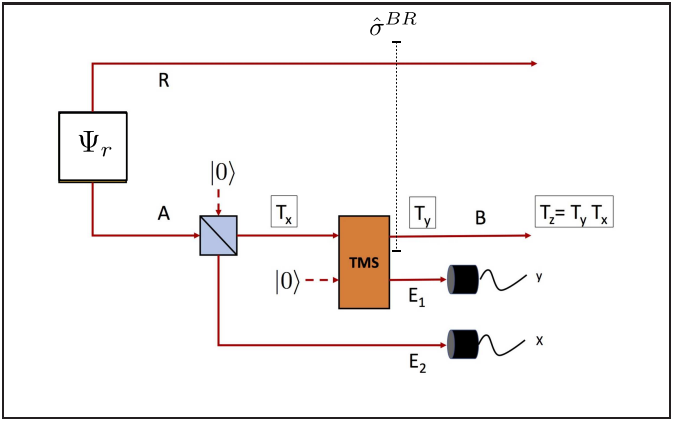}}
\caption{Coherent representation of Kraus operators of noisy attenuator and amplifier channels obtained through the concatenation of suitably chosen quantum-limited attenuators and amplifiers. Using the Stinespring dilation of the quantum-limited attenuator and amplifier, one obtained a purification of the state $\hat{\sigma}^{BR}$ corresponding to the one-sided action of the noisy channel action on a two-mode squeezed state $|\Psi_r \rangle$, and $E_1$ and $E_2$ are the purification modes of $\hat{\sigma}^{BR}$. We then measure the auxiliary modes of the quantum-limited channels in the Fock basis individually. Conditioned on the outcome measurement $y$ and $x$ of the environment modes $E_1$ and $E_2$, and in the limit of arbitrary large squeezing, one obtains $T_z$ as a product of Kraus operators $T_y$ and $T_x$. Here TMS stands for the two-mode squeeze operator.}
\label{Fig:figure1}
\end{figure}
\begin{figure}
\fbox{\includegraphics[width=\columnwidth]{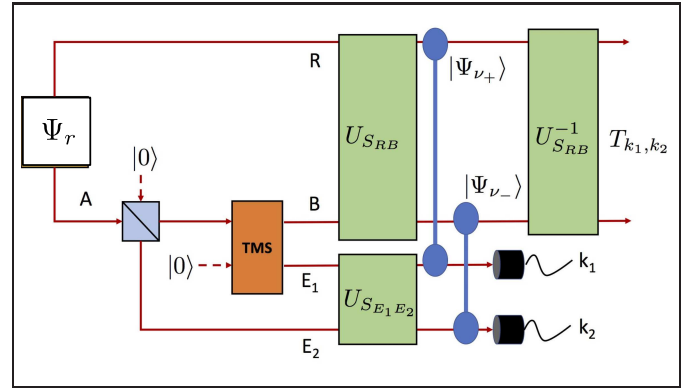}}
\caption{Coherent representation of Kraus operators obtained through our method outlined in Sec. \ref{tech}. One mode of an initial two-mode squeezed state $\ket{\psi_r}$ is sent through the noisy attenuator or amplifier channel obtained by concatenation of the quantum-limited amplifier and attenuator channel, with the channel represented in its respective Stinespring dilation. By acting locally by unitaries $U_{S_{RB}}, U_{S_{E_1E_2}}$ on a bipartite system with one subsystem represented by the output of the channel B and the purifying system R, and the other system by the auxiliary modes $E_1$ and $E_2$, the total output four-mode pure state is taken to a product of two-mode squeezed states $\ket{\psi_{\nu_+}} \otimes \ket{\psi_{\nu_-}}$; this is possible since all the involved states are Gaussian states. Conditioned on the measurement output $k_1,k_2$ through a suitable POVM on the auxiliary modes $E_1,E_2$, and undoing the local unitary $U_{S_{RB}}^{-1}$ on the  output and purifying system $RB$, we obtain the corresponding double-indexed Kraus operators $T_{k_1,k_2}$ (in the limit of arbitrary large squeezing). This method is to be compared with the scheme mentioned in Fig. \ref{Fig:figure1} where there is no joint processing of the auxiliary modes which are in fact measured individually.  }
\label{Fig:figure2}
\end{figure}
As shown in Fig. \ref{fig1}, using the Stinespring representation, the Kraus operators of the quantum-limited channels are obtained
by measuring its respective environment modes in the appropriate basis (Fock states or coherent states). As detailed in \cite{gaussch}, one approach to obtain a set of Kraus operators for noisy amplifiers and attenuator channels is by concatenation of two quantum-limited ones.
Therefore, as shown in Fig. \ref{Fig:figure1}, one can construct the Kraus operators
$T_z$ of the noisy channels [of Eq.\,(9.1) of (\cite{gaussch})], by composing the
Kraus operators of the quantum-limited attenuator and
amplifier channels, i.e., $T_z=T_xT_y$, where $z$ is discrete and double indexed.

We have seen in Sec. \ref{tech} (Fig. \ref{fig2}) that there exist a unitary $U_{S_{RB}}$ that transforms the Choi state
into a tensor product of two thermal states, which allows us to find a decomposition of $\hat{\sigma}(r)$ depending on the decomposition of the thermal states into 
Fock states or coherent states. 
The quantum state of the modes $R,B,E_1,E_2$ in Fig. \ref{Fig:figure2} is a pure state and can be seen as a bipartite system with $RB$ as one subsystem and $E_1E_2$ the other. 
It is known that all bipartite pure Gaussian
states can be mapped into a tensor product of two-mode squeezed vacuum states
by local Gaussian unitary operations\,\cite{Giedke2003}. So we know that there exists a pair of
symplectic operations $S_{E_1E_2}$ and $S_{RB}$, where $S_{RB}$ realizes the symplectic diagonalization
of modes $R$ and $B$ (during step 2 of the earlier derivation), that transforms the global
pure state into two pairs of two-mode squeezed vacuum. This is a purification of the
state of Eq.\,(\ref{k9}) one obtains after step 2 (of Fig.\,\ref{fig2}). Therefore by properly selecting
the symplectic operations $S_{E_1E_2}$ that contributes partly to the transformation of the global
pure state into two pairs of two-mode squeezed vacuum, and applying the
correct rank-one POVM, one generates the Kraus operators obtained in the previous section instead of those of\,\cite{gaussch}. \\ 

\noindent
{\bf Remark.} As hinted in the previous remark in Sec. \ref{naa}, there is a third straightforward method of obtaining Kraus operators for noisy channels. 
This involves composing the Kraus operators of the quantum-limited channel with Heisenberg-Weyl 
displacement operators with an appropriate Gaussian weight, which are the Kraus operators of the additive Gaussian classical noise channel.\hfill $\blacksquare$

\section{Singular channel ${\cal A}_{2}(\alpha)$ \label{secsing}}
The singular channels ${\cal A}_{2}(\alpha)$
are characterized (in the canonical form) by $(X,Y)= ((1\!\!1 + \sigma_3)/2, \alpha 1\!\!1)$, i.e. the transfer matrix $X$ is singular, and $\alpha\geq1$.
During step 1 of our procedure outlined in Section \ref{tech}, the output state variance matrix reads
\begin{align}\label{s1}
V^s(r) = \left(
\begin{matrix}
\cosh{2r}+\alpha  & 0 & \sinh{2r} & 0 \\
0 & \alpha & 0 & 0\\
\sinh{2r} & 0 & \cosh{2r} & 0\\
0 & 0 & 0 & \cosh{2r}
\end{matrix}
\right),
\end{align}
where the superscript `$s$' denotes a singular channel. Since the off-diagonal block of $V^s_{\rm{out}}(r)$ is singular, the corresponding two-mode Gaussian state is manifestly separable by the Lemma of\,\cite{simon00}, and this is true independent of the numerical value of squeezing parameter $r$. The channel is thus entanglement breaking.

In the case of a quantum-limited singular channel we have $\alpha=1$, and during
step 2 of the derivation of the Kraus operators
we perform the local symplectic (scaling) transformation
$S_{1}(r)={\rm diag}(1,1,(\cosh{2r})^{1/2},(\cosh{2r})^{-1/2}) \in {\rm Sp}(4,{\mathbb R})$ to render
the $2 \times 2$ `momentum' block of
${V}^s(r)$ an identity matrix, resulting in
\begin{align}
V^{(1)}(r)=
\left[\begin{matrix}
\cosh{2r}+1  & 0 & \sinh{2r}\,(\cosh{2r})^{1/2} & 0 \\
0 & 1 & 0 & 0\\
\sinh{2r}\,(\cosh{2r})^{1/2} & 0 &\cosh^2{2r} & 0\\
0 & 0 & 0 & 1
\end{matrix}
\right].
\end{align}
It is clear that this can be
diagonalized using a beam splitter-type rotation in ${\rm Sp}(4,{\mathbb R})$ (equal rotation in position and momentum planes), leading to the canonical form
\begin{align}
V^s_{\rm can}(r)&={\rm diag}(1+\lambda, 1,1,1),\nonumber\\
\lambda&=\sinh^2{2r}+\cosh{2r},
\label{s2}
\end{align}
the diagonalizing symplectic rotation being
\begin{align}
 S(\theta(r))&=
\left( \begin{array}{cc}
  \cos{\theta} 1\!\!1_2 & - \sin{\theta} 1\!\!1_2 \\
 \sin{\theta} 1\!\!1_2 & \cos{\theta} 1\!\!1_2
\end{array} \right), \nonumber \\
\cos{\theta}&=\frac{\sqrt{\cosh{2r}}}{\sqrt{\lambda}},~~ \sin{\theta}=\frac{\sinh{2r}}{\sqrt{\lambda}}.
\label{sings1}
\end{align}
In step 3 we decompose
$\hat{\sigma}^s_{\rm can}(r)$ into a Gaussian mixture of coherent states along the `position' quadrature.
The Gaussian state corresponding to ${V}^s_{\rm can}(r)$ in Eq. \eqref{s2}  reads as
\begin{align}\label{s6}
\hat{\sigma}^s_{\rm can}(r) &=
 \int \frac{dx^{\,'}}{\sqrt{\pi\lambda}} \,\exp[-(x^{\,'})^2/\lambda]\, \nonumber\\
&~~~~~~~~~~~~~~\times D(x^{\,'})\proj{0}{D}^{\dagger}(x^{\,'})\otimes\proj{0}.
\end{align}
In step 4 we perform on $\hat{\sigma}^s_{\rm can}(r)$ the inverse of the symplectic diagonalization applied in the two earlier stages. First, we undo
the beam splitter rotation of Eq. \eqref{sings1} so that the single-mode coherent state $D(x^{\,'})\ket{0}\otimes\ket{0}$
is mapped to a two-mode coherent state\,:
\begin{align}
D(x^{\,'})\ket{0}\otimes\ket{0}\to
D\left(x^{\,'}\,\cos{\theta} \right)\ket{0}\otimes
D\left(x^{\,'}\,\sin{\theta} \right)\ket{0}.
\end{align}
Then we undo the local squeeze transformation $S_{1}(r)$,
so that the pure states in the decomposition of $\hat{\sigma}^s(r)$ attain, for each $x^{\,'}$, the form
\begin{align}
&D\left(x^{\,'}\,\cos{\theta} \right)\ket{0} \otimes
U[S_{1}(r)]^{\dagger} D\left(x^{\,'}\,\sin{\theta} \right) \,\ket{0}\nonumber\\
&=D\left(x^{\,'}\,\cos{\theta} \right)\ket{0} \otimes
D\left(\frac{x^{\,'}\sin{\theta}}{\sqrt{\cosh{2r}}} \right) U[S_{1}(r)]^{\dagger}\ket{0}.
\end{align}
We have
\begin{widetext}
\begin{align}
\hat{\sigma}^s(r) =  \int \frac{dx^{\,'}}{\sqrt{\pi\lambda}} \,\exp[-[x^{\,'}]^2/\lambda]\,
D\left[x^{\,'}\,\cos{\theta} \right]\proj{0}\,D\left[x^{\,'}\,\cos{\theta} \right]^{\dagger}  \otimes
D\left[\frac{x^{\,'}\sin{\theta}}{\sqrt{\cosh{2r}}} \right] U[S_{1}(r)]^{\dagger}\proj{0} \,U[S_{1}(r)]\, D\left[\frac{x^{\,'}\sin{\theta}}{\sqrt{\cosh{2r}}} \right]^{\dagger}.
\end{align}
Let us now denote $x^{\,'} \cos{\theta}$ by $x$, so that
\begin{align}
&\hat{\sigma}^s(r) =  \int\frac{dx}{\cos{\theta}\,\sqrt{\pi\lambda}} \,\exp[-[x]^2 /\lambda\,\cos^2{\theta}]\,
  D(x)\proj{0}\,D(x)^{\dagger}
\otimes
D\left[\frac{x\tan{\theta}}{\sqrt{\cosh{2r}}} \right] U[S_{1}(r)]^{\dagger}\proj{0} \,U[S_{1}(r)]\, D\left[\frac{x\tan{\theta}}{\sqrt{\cosh{2r}}} \right]^{\dagger}.
\end{align}
\end{widetext}

Substituting for $\theta$ from Eq.\,\eqref{sings1} we obtain
\begin{align}\label{s10}
\hat{\sigma}^s = {\rm sech}^2{r}\,\int dx \, |T_x^s(r)\rangle \langle T_x^s(r)|,
\end{align}
where
\begin{align}
|T_x^s(r)\rangle &= \sqrt{N_s(r)} \,\exp\left[-\frac{x^2}{2 \cosh{2r}}\right]\, \nonumber\\ &\times D(x)|0\rangle \otimes D(x\tanh{2r})\,U[S_{1}(r)]^{\dagger} |0\rangle,
\label{krausket}
\end{align}
and
\begin{align}
N_s = \frac{\cosh^2{r}}{\sqrt{\pi\cosh{2r}}}.
\label{ns}
\end{align}
We can now read off the Kraus operator $T_x^s(r)$ from the bipartite vector $|T_x^s(r)\rangle$ in Eq. \eqref{krausket} as
\begin{align}
T_x^s(r) &= \sqrt{N_s(r)} \,\exp\left[-\frac{x^2}{2\cosh{2r}}\right]\, \nonumber\\ &~~~~ \times D(x) \proj{0} U[S_{1}(r)]\,D(x\tanh{2r})^{\dagger}.
\label{ksr}
\end{align}
The completely positive map ${\cal A}_2(r)$ deduced by the action of the noiseless ${\cal A}_2$ channel on $|\Psi_r\rangle$ is not trace-preserving, as seen by evaluating the operator integral 
\begin{widetext}
\begin{align}
&\int dx^{\,'}\, T_{x^{\,'}}^s(r)^{\dagger}\, T_{x^{\,'}}^s(r)
=\int dx^{\,'}\, N_s(r) \exp[-(x^{\,'})^2/\cosh{2r}]\, D(x^{\,'}\,\tanh{2r}) \,U[S_1(r)]^{\dagger} \proj{0} U[S_1(r)]\,D(x^{\,'}\,\tanh{2r})^{\dagger}\nonumber\\
&=\int dx \,\frac{\cosh^2{r}}{(\pi \cosh{2r})^{1/2}\tanh{2r}} \, \exp\left[-\frac{x^2}{(\cosh{2r}\,\tanh^2{2r})}\right]\, D(x)\,U[S_1(r)]^{\dagger} \proj{0} U[S_1(r)]\, D(x)^{\dagger},
\label{singtp}
\end{align}
\end{widetext}
where $x = x^{\,'} \,\tanh{2r}$.

It is easy to see that this operator integral is a thermal state with variance matrix $\cosh{2r} \,1\!\!1_2$, except for the overall factor of $\cosh^2{r}$.
The squeezing operation $U[S_1(r)]^{\dag}$ maps the vacuum state to a (single-mode) squeezed state with  variance matrix $V_1(r) = {\rm diag}\,\left(({\cosh{2r})^{-1},\,\cosh{2r}}\right)$. Then the convex sum action of the displacement operators, with Gaussian weight factor, maps $V_1(r) \to V_2(r) = {\rm diag}\,({(\cosh{2r})^{-1} + \cosh{2r}\,\tanh^2{2r} ,\,\cosh{2r}})$, which evaluates to $\cosh{2r}\, 1\!\!1$.

Expanding this thermal state in the Fock basis, and accounting for the multiplicative factor of $\cosh^2{r}$ in Eq.\,\eqref{singtp}, we get
\begin{align}
\int dx\, T_x^s(r)^{\dagger}\, T_x^s(r) = \sum_{n=0}^{\infty} \tanh^{2n}{r} \,\proj{n}.
\end{align}
Going back to Eq.\,\eqref{ksr}, we write the squeezed vacuum state as a superposition of Fock states\,\cite{yuen,lvovsky}, i.e., 
\begin{align}
U[S_1(r)]^{\dagger} |0\rangle &=  \sqrt{{\rm sech}\,{\xi_r}} \sum_{n=0}^{\infty} \, \frac{\sqrt{2n!}}{n!} \,\left(-\frac{\tanh{\xi_r}}{2}\right)^{n} |2n\rangle \nonumber\\ & \equiv \sqrt{{\rm sech}\,{\xi_r}}\, |\xi_r\rangle,
\label{xir}
\end{align}
where $\exp[\xi_r] = \sqrt{{\rm sech}\,{2r}}$. The exponential Gaussian weight factor in  Eq.\,\eqref{ksr} grows rapidly flat with increasing $r$, and goes to a constant as $r \to \infty$. Further, $\tanh{r} \to 1$ in this limit, so we have
\begin{align}
\sqrt{N_s(r)}\, U[S_1(r)]^{\dagger} |0 \rangle &= \left[\frac{\cosh^2{r}}{\sqrt{\cosh{2r}}}~ {\rm sech}\,{\xi_r}\right]^{1/2}  \left[ \frac{1}{\pi^{1/4}} |\xi_r\rangle \right]\nonumber\\
& \,\to |0\rangle_{\rm pos},
\end{align}
the square-root factor going to unity in this limit, and $|\cdot\rangle_{\rm pos}$ is a position eigenket. Here we have used the fact that the scalar $\tanh{\xi_r}$ in the Fock basis expansion of $|\xi_r \rangle$ in Eq.\,\eqref{xir} tends to unity and the remaining factors are identified as the Fock state representation of position ket $\ket{0}_{\rm pos.}$ located at zero\,\cite{qmbook}.  Note that ${\rm tanh} 2r \to 1$ as $r \to \infty$. Collecting the above facts, we have
\begin{align}
\lim_{r \to \infty} T_x^s(r)  = \left( |x/\sqrt{2}\rangle_{\rm coh}\right)\, {}_{\rm pos}\langle x| \equiv A_x,
\end{align}
where $|x/\sqrt{2}\rangle_{\rm coh}$ is a coherent state. In the limit $r\to \infty$ we recover the condition
\begin{align}
\lim_{r \to \infty} \int dx\, T_x^s(r)^{\dagger}\, T_x^s(r)  = \int dx A_x^{\dagger} A_x = 1\!\!1.
\end{align}
It is clear that the Kraus operator $A_x$ acting on a state $|\psi\rangle$ {\em computes the component of $|\psi\rangle$ along position eigenket $|x\rangle_{\rm pos}$, and outputs the coherent state $|x/\sqrt{2}\rangle_{\rm coh}$ with amplitude ${}_{\rm pos}\langle x|\psi\rangle$}, resulting in a measurement-preparation interpretation of the channel. \\

\noindent
{\bf Remark.}
This procedure can be readily extended to obtain rank-one Kraus operators
for the noisy singular channel ${\cal A}_2(\alpha)$, $\alpha >1$, in a manner similar to the
earlier cases.
 In step 3 of the procedure above, we find in the noisy case $\hat{\sigma}_{\rm can}(r)$ to be
a product of two thermal states of nonvanishing temperature, and we expand them both in the coherent state
basis. In the limit $r \rightarrow \infty$ we are left with
a Gaussian coherent state decomposition
for $\hat{\sigma}(r)$. The Kraus operators are now
written by flipping the second ket to a bra of the product
coherent state composing $\lim\limits_{r \to \infty} \hat{\sigma}(r)$.
Note that now there is an additional multiplicative Gaussian factor in each of the
rank-one Kraus operators.\,\hfill $\blacksquare$\\

The final example of the single-quadrature noise channel ${\cal B}_1$  is not entanglement breaking. Nevertheless, its Kraus operators can be obtained in a manner analogous
to that of ${\cal A}_2$ as briefly outlined below.

\subsection{Single quadrature noise ${\cal B}_1$}
The single quadrature noise is the final of the Gaussian channels that we consider. We mention the basic steps which leads to the operator-sum representation for these channels. 
The single quadrature noise channel is specified by matrices $X=1\!\!1$ and $Y=\alpha(1\!\!1 + \sigma_3)/2$.
Under its one-sided action, $V_{\rm in}(r) \to V(r)= V_{\rm in}(r) + {\rm diag}(\alpha,0,0,0)$.
During step 2 we first apply a squeeze transformation, i.e., $V(r)
\to V_{\rm in}(r)^{-\frac{1}{2}}\, V(r)\, V_{\rm in}(r)^{-\frac{1}{2}}$, and then follow
it by a suitable rotation to obtain $V_{\rm can}(r)$. The corresponding $\hat{\sigma}_{\rm can}(r)$
is as in Eq.\,(\ref{s6}) with the appropriate $\lambda$. That is, $\hat{\sigma}_{\rm can}(r)$
is the ground state displaced along the $x$ quadrature of the first mode with a Gaussian
distribution times the ground state of the second mode.
It is a simple exercise to check that undoing the rotation and
two-mode squeezing on $\hat{\sigma}_{\rm can}(r)$ yields a mixture of $|\Psi(r) \rangle$
displaced along the $x$ quadrature of the first mode with a Gaussian distribution.
In the large $r$ limit, with appropriate relabeling of variables, we recover the displacement
operators on a single quadrature weighted by a Gaussian function to be the Kraus
operators as expected\,\cite{gaussch}.


\end{document}